\documentclass[prd,aps,
floats,letterpaper,floatfix,
groupedaddress,superscriptaddress
,nofootinbib
,eqsecnum
,twocolumn
]{revtex4}
\usepackage{amsmath}
\usepackage{amsfonts}
\usepackage{bm}
\usepackage{graphicx}
\usepackage[usenames]{color}
  \newcommand{\apjl}{Astrophys.\ J.\ Lett.}

 \newcommand{\jcap}{J.\ Cosmol.\ Astropart.\ Phys.}

\allowdisplaybreaks
\def\eae{Einstein-{\AE}ther }

\begin{document}

\title{Compact binary systems in Einstein-{\AE}ther gravity. II. Radiation reaction to 2.5 post-Newtonian order}
\author{
Fatemeh Taherasghari} \email{fatemet@illinois.edu}
\affiliation{Ilinois Center for Advanced Studies of the Universe \& Department of Physics, University of Illinois at Urbana-Champaign, Urbana, Illinois 61801, USA}

\author{
Clifford M.~Will} \email{cmw@phys.ufl.edu}
\affiliation{Department of Physics, University of Florida, Gainesville, Florida 32611, USA}
\affiliation{GReCO, Institut d'Astrophysique de Paris, CNRS,\\ 
Sorbonne Universit\'e, 98 bis Bd.\ Arago, 75014 Paris, France}

\date{\today}

\begin{abstract}
We obtain the equations of motion for compact binary systems (black holes or neutron stars) in an alternative theory of gravity known as \eae theory, which supplements the standard spacetime metric with a timelike four-vector (the {\AE}ther field) that is constrained to have unit norm.  The equations make use of solutions obtained in Paper I for the gravitational and {\AE}ther field potentials within the near zone of the system, evaluated to 2.5 post-Newtonian (PN) order ($O(v/c)^5$ beyond Newtonian gravity), sufficient to obtain the effects of gravitational radiation reaction to the same order as the quadrupole approximation of general relativity.  Those potentials were derived by applying the post-Minkowskian method to the field equations of the theory.
Using a modified geodesic equation that is a consequence of the effects of the interaction between the {\AE}ther field and the internal strong-gravity fields of the compact bodies, we obtain explicit equations of motion in terms of the positions and velocities of the bodies, focussing on the radiation-reaction terms that contribute at 1.5PN and 2.5PN orders.  We obtain the rate of energy loss by the system, including the effects of dipole gravitational radiation (conventionally denoted $-1$PN order) and the analogue of quadrupole radiation (denoted $0$PN order).  We find significant disagreements with published results, based on calculating the energy flux in the far zone using a ``Noether current'' construction.
\end{abstract}

\maketitle

\section{Introduction}
\label{sec:intro}

This is the second in a series of papers that explore an alternative gravitational theory known as \eae gravity.  This theory was devised by Jacobson and collaborators
\cite{2001PhRvD..64b4028J,2002cls..conf..331M,2004PhRvD..70b4003J,2004PhRvD..69f4005E,2006PhRvD..73f4015F} as a model for studying effective Lorentz symmetry violation in gravitational interactions.  This theory generated substantial interest in part because it was a natural extension of the approach that led to scalar-tensor theories:  start with general relativity and add an additional field that does not couple directly with matter (so as to preserve the principle of equivalence).  In this case, the added field was a timelike four-vector field instead of a scalar field.  Numerous variants of the theory appeared, including TeVeS \cite{2004PhRvD..70h3509B}, Khronometric theory 
\cite{2009PhRvD..79h4008H}, and other scalar-tensor-vector theories  \cite{2006JCAP...03..004M,2008PhRvD..77l3502S,2009CQGra..26n3001S}.  At the lowest post-Newtonian (PN) order, the parametrized post-Newtonian (PPN) parameters of \eae theory were calculated by Foster and Jacobson \cite{2006PhRvD..73f4015F}; the values were identical to those of general relativity, except for the ``preferred-frame'' parameters, $\alpha_1$ and $\alpha_2$, which could be non-zero, thus generating Lorentz violating effects that could be tested by experiments in the solar system and beyond.   In time a substantial literature developed (more than 200 papers as of 2025), examining everything from cosmology to black holes in \eae gravity.  

With the discovery of binary pulsars and  the advent of gravitational-wave astronomy,  many authors addressed the predictions of \eae theory for gravitational waves  
\cite{2006PhRvD..73j4012F,
2007PhRvD..75l9904F,
2007PhRvD..76h4033F,
2014PhRvD..89h4067Y,
2018PhRvD..97h4040G,
2018PhRvD..97l4023O,
2018Univ....4...84H,2018PhRvD..98l4015S,
2019PhRvD..99b3010L,
2020PhRvD.101d4002Z,
2021CQGra..38s5003G,
2023PhRvD.108j4053S}.
However, none of this work carried the calculations beyond the first post-Newtonian (1PN) order in the equations of motion for binary systems (apart from a partial calculation by Xie and Huang \cite{2008PhRvD..77l4049X}), or beyond the equivalent of $0$PN order (corresponding to the quadrupole formula of general relativity) for the gravitational waveform or energy loss rate.  

By contrast, calculations to high PN orders have been carried out in general relativity (GR) \cite{2024LRR....27....4B} and scalar-tensor gravity 
\cite{2013PhRvD..87h4070M,2014PhRvD..89h4014L,2015PhRvD..91h4027L,2018PhRvD..98d4004B,
2022JCAP...08..008B}, in part responding to the requirements of data analysis at the ground-based gravitational wave interferometers.  These calculations were based on the Landau-Lifshitz or post-Minkowskian formulation of the field equations for these theories, which rewrites the field equations in the form of flat-spacetime wave equations for metric (and scalar) perturbations whose source is the matter energy momentum tensor plus field contributions that are quadratic and higher in the small perturbative quantities (see \cite{PW2014} for a pedagogical treatment).

Accordingly we endeavored to apply this highly successful method to \eae theory, in hopes of obtaining similarly high-order PN results, beginning with \cite{2023PhRvD.108l4026T} (hereafter referred to as Paper I).   In light of the extremely tight constraint on the difference in speed of gravitational waves compared to electromagnetic waves obtained from observations of the gravitational-waves and gamma ray bursts from a binary neutron star merger in 2017 
\cite{2017PhRvL.119p1101A,2017ApJ...848L..13A}, we restricted attention to a subset of \eae theories whose parameters satisfy the constraint $c_3 +c_1 = 0$.  

We found that a naive application of the post-Minkowskian method failed, producing a set of coupled equations involving double spatial and time derivatives (plus mixed spatial-time derivatives) of the fields at linear order, along with matter sources and non-linear field contributions.   This was due in part to the inherent complexity of working with a four-vector field rather than a scalar field, and in part to the constraint that  the {\AE}ther field have unit norm, which is a key aspect of the theory.
We then found a way to redefine the perturbative metric and {\AE}ther field variables that decoupled the equations and produced a flat-spacetime wave equation for each field in a form that was parallel to that in GR, with right-hand sides consisting of matter sources plus terms quadratic and higher order in the small perturbative fields.   Following the method of Direct Integration of the Relaxed  Equations (DIRE) \cite{1996PhRvD..54.4813W,2000PhRvD..62l4015P}, we obtained PN solutions for the fields within the near zone of slowly moving systems up to 2.5PN order, corresponding to the quadrupole-formula order for gravitational radiation damping within GR.  Since our goal was to treat compact bodies,  we made the conventional assumption \cite{1975ApJ...196L..59E} that the {\AE}ther field interacts indirectly with the bodies via coupling to their internal gravitational fields (because \eae theory is a metric theory, 
the {\AE}ther field does not couple directly to matter fields).  

In this paper we assume that the source is a binary system of compact objects, and, using our solutions from Paper I, we obtain the equations of motion through 2.5PN order.   The equations are expressed in terms of a number of ``sensitivity parameters'' for each body, that encode how the mass of the body changes when the ambient {\AE}ther field changes.
At Newtonian and 1PN order, our equations agree completely with earlier results.  The 2PN contributions will be discussed in a future publication.  Here we focus on the 1.5PN terms, which represent the effects of gravitational radiation associated with the dipole moment of the {\AE}ther field, and the 2.5PN terms which include conventional quadrupole and higher contributions analogous to those in GR, plus PN corrections of the 1.5PN dipole terms.  Using these equations of motion through 1PN order, we obtain the total energy of the system, which is conserved through that order.  We then calculate its rate of change using the 1.5PN and 2.5PN terms in the equations of motion, obtaining the $-1$PN and $0$PN contributions to $dE/dt$.  (For quantities like energy flux and gravitational waveforms, it is conventional to begin the PN numbering with zero, for the analogue of GR quadrupole radiation, where it is the leading contribution, then $1$PN for the first PN corrections and so on, and to use $-1$PN for dipole energy loss, which is of order $1/v^2$ larger than $0$PN energy loss).  

However, already at $-1$PN order in $dE/dt$, our result disagrees with earlier work \cite{2006PhRvD..73j4012F,2007PhRvD..76h4033F,2014PhRvD..89h4067Y}.  These papers calculate the energy flux in the far zone using a construction known as the ``Noether current'' combined with a linearized approximation for obtaining the far-zone fields.  In the linear approximation of \eae theory in vacuum, the equations yield three different speeds of propagation for the fields, a speed of unity (a consequence of the constraint $c_3 = -c_1$) for transverse-traceless metric perturbations, a speed of $v_T$ for transverse {\AE}ther waves (with no metric perturbations) and a speed of $v_L$ for longitudinal waves that couple {\AE}ther and metric perturbations
(see Eq.\ (\ref{eq:wavespeeds2}) for definitions).  Our expression for $dE/dt_{-1PN}$ depends only on $v_T$, while the result from \cite{2006PhRvD..73j4012F,2007PhRvD..76h4033F,2014PhRvD..89h4067Y} depends on both $v_T$ and $v_L$.   For $dE/dt_{0PN}$ the disagreement even larger; we believe that a part of this is the result of the linearized approximation used by these authors.  Even in GR, this assumption is formally incorrect, but leads to an answer of the correct form (the ``quadrupole formula'') by what amounts to a fluke; in many alternative theories of gravity it is simply not correct \cite{1977ApJ...214..826W}. 

The remainder of this paper gives details.  Section \ref{sec:EOM1} describes the derivation of the equations of motion for compact bodies through 1.5PN order plus 2.5PN order using our solutions from Paper I, expressed formally in terms of gravitational potentials and multipole moments and their derivatives.  In Sec.\ \ref{sec:radreaction} we specialize to binary systems, and express the equations in terms of positions and velocities of the bodies, and in Sec.\  \ref{sec:energyloss} we calculate the energy loss rate.   Section \ref{sec:discussion} discusses the results, notably the disagreement with other published work.  In a series of Appendices, we obtain expressions for the multipole moments for binary systems (including the PN corrections to the {\AE}ther dipole moment); derive the total energy and momentum of the system to PN order, along with the PN-corrected transformation between individual velocities and the relative velocity; display the very complicated coefficients that appear in the $0$PN energy loss rate; and correct some typos that appeared in Paper I.

We use units in which the speed of light $c$ is unity, and the spacetime metric has the signature $(-,+,+,+)$; Greek indices denote spacetime components and Roman indices denote spatial components; parentheses (square brackets) around groups of indices denote symmetrization (antisymmetrization).

\section{Compact body equations of motion to 2.5 post-Newtonian order}
\label {sec:EOM1}

\subsection{Quick review of Paper I}
\label{sec:review}

In Paper I \cite{2023PhRvD.108l4026T}, we used the method of Direct Integration of the Relaxed Equations (DIRE), adapted to \eae theory to get the metric in the near-zone of a system in a post-Newtonian expansion, 
\begin{align}
g_{00} &= -1 +   \frac{\epsilon}{2} \tilde{N} 
+  \frac{\epsilon^2}{8}  \left ( 4\tilde{B} - 3 \tilde{N}^2 \right )
\nonumber \\
& \quad
 +  \frac{\epsilon^3}{16} \left ({5} \tilde{N}^3 - {4} \tilde{N}\tilde{B} +8 \tilde{K}^j \tilde{K}^j  \right )
 +O(\epsilon^4) \,, 
\nonumber\\
g_{0j} &= -  \epsilon^{3/2} \tilde{K}^j  +   \frac{\epsilon^{5/2}}{2} \tilde{N}  \tilde{K}^j  +O(\epsilon^{7/2}) \,, 
\nonumber \\
g_{jk} &= \delta^{jk} \left \{ 1+  \frac{\epsilon}{2} \tilde{N} -  \frac{\epsilon^2}{8}  \left ( \tilde{N}^2 + 4 \tilde{B}  \right ) \right \} 
\nonumber \\
&
\quad 
+  \epsilon^2 \tilde{B}^{jk} 
 +O(\epsilon^3) \,, 
\nonumber \\
(-g) &= 1+  \epsilon\tilde{N}  - \epsilon^2 \tilde{B} + O(\epsilon^3) \,,
\label{eq:metric}
\end{align}
where $\epsilon \sim v^2 \sim Gm/r$ is the standard PN expansion parameter.
The \AE ther field is given by $(K_{\rm ae}^0, K_{\rm ae}^j)$, where the constraint 
$K_{\rm ae}^{\mu}K_{{\rm ae}\mu}= -1$ implies that 
\begin{equation}
K_{\rm ae}^0 = 1 + \frac{1}{4} \epsilon \tilde{N} + \frac{1}{4} \epsilon^2 \left ( \tilde{B} + \frac{3}{8} \tilde{N}^2 \right ) + O(\epsilon^3) \,.
\label{eq:K0}
\end{equation}

A crucial step in Paper I was a change of variables from the metric fields $\tilde{N}$, $\tilde{B}$, $\tilde{B}^{jk}$, $\tilde{K}^j$ and $\tilde{K}^j_{\rm ae}$ to a new set of fields $N$, $B$, ${B}^{jk}$, ${K}^j$ and ${K}^j_{\rm ae}$ etc (see Paper I, Eqs.\ (4.7)) that allowed us to express the field equations in a ``post-Minkowskian'' form in which each of the new fields satisfies a flat spacetime wave equation, with a source consisting of the matter energy momentum densities plus field contributions that are quadratic and higher in the fields themselves, the same structure as in the ``relaxed Einstein equations'' of general relativity.   These wave equations had the form
 \begin{align}
  \left (1-\frac{1}{2}c_{14}\right) \Box N &= -16\pi G_0 \tau^{00} + O(\rho \epsilon^3) \,,
\nonumber \\
\Box K^j &= -16\pi G_0 \tau^{0j}  + O(\rho \epsilon^{5/2}) \,,
\nonumber  \\
\Box B^{jk} &= -16\pi G_0 \tau^{jk}  + O(\rho \epsilon^{2}) \,,
\nonumber  \\
\Box B &= -16\pi G_0  \tau^{kk}+ O(\rho \epsilon^{3}) \,,
\nonumber  \\
c_1 \Box^* K_{\rm ae}^j &= 8 \pi G_0 \tau_{\rm  ae}^j  + O(\rho \epsilon^{5/2}) \,,
\label{eq:fieldeqsnew}
\end{align}
where $\Box^* \equiv \nabla^2 - v_T^{-2} \partial_0^2$,  and where
\begin{align}
\tau^{\mu\nu} &\equiv (-g) T_T^{\mu\nu} + (16\pi G_0)^{-1} \Lambda_T^{\mu\nu} \,,
\nonumber \\
\tau_{\rm ae}^j & \equiv T_{{\rm  ae} \, T}^j  + (8 \pi G_0)^{-1} \Lambda^j_{\rm ae} \,.
\label{eq:tau}
\end{align}
From the linearized \eae equations in vacuum, it is known that there are solutions in the wave zone with three characteristic speeds, unity for tensor gravitational  waves (a consequence of our post-2017 constraint $c_1 + c_3 =0$), $v_T$ for transverse {\AE}ther waves, and $v_L$ for longitudinal waves that couple {\AE}ther and metric fields (see Appendix A of Paper I for a derivation), where
\begin{equation}
v_T^2 = \frac{c_1}{c_{14}}   \,, \quad v_L^2 = \frac{(2-c_{14}) c_2}{(2+3c_2) c_{14}} \,.
\label{eq:wavespeeds2}
\end{equation}
Notice that in Eqs.\ (\ref{eq:fieldeqsnew}), in terms of our new variables, the wave equations for the metric fields involves the speed unity and the wave equation for the {\AE}ther field involves $v_T$.  There is no wave equation involving the speed $v_L$.  

Focusing on the near zone of the system, we then obtained solutions for those fields through 2.5PN order in terms of instantaneous Poisson-like  potentials and their generalizations, of the form, 
\begin{align}
    P(f)&\equiv\frac{1}{4\pi}\int_{\mathcal{M}}\frac{f(t,\bm{x'})}{|\bm{x}-\bm{x'}|}d^3x',  \;\;\;\nabla^2 P(f)=-f \,,
    \nonumber\\
    \Sigma(f)&\equiv \int_{\mathcal{M}} \frac{\sigma(t,\bm{x'})f(t,\bm{x'})}{|\bm{x}-\bm{x'}|}d^3x'=P(4\pi \sigma f) \,,
\end{align}
\noindent where the integration is confined to the near-zone $\mathcal{M}$, a sphere of radius of order one gravitational wavelength surrounding the system. The potentials of Paper I were expressed in terms of the specific source densities given by
\begin{align}
    \sigma &\equiv T^{00}_{T} + T^{ii}_{T},  &\sigma^i &\equiv T^{0i}_{T}  \,,\nonumber \\
    \sigma^{jk} &\equiv T^{jk}_{T}, & \sigma^j_{\text{ae}} &\equiv T^j_{T\text{ae}} \,,
\end{align}
\noindent 
where $T^{\alpha\beta}_{T}$ are a combination of the matter energy-momentum tensor  $T^{\mu\nu}$ and the {\AE}ther energy-momentum vector  $T^\mu_{\rm ae}$, defined in Paper I, Eqs.\ (3.2).   

\subsection{A modified geodesic equation}
\label{sec:modifiedgeodesic}

To obtain the equations of motion for non-spinning compact bodies in \eae theory, we begin by finding a general ``modified geodesic equation'' for compact bodies in the theory.  The modifications arise from the fact that we are treating self-gravitating bodies, whose internal structure and mass could depend on external values of the {\AE}ther field.  Using the Bianchi identity applied to the generally covariant matter action for self-gravitating bodies in a broad class of alternative theories of gravity, we obtained general forms for such modified geodesic equations (see \cite{2022PhRvD.106f4021T}, Eq.\ (3.15)).  For \eae theory, the equations reduced to 
    \begin{align}
    {T^\nu}_{\alpha;\nu}=-T_{\rm ae\nu} K^\nu_{\rm ae;\alpha} -\left(K^\nu_{\rm ae} T_{\rm ae \alpha}\right)_{;\nu},
    \label{eq:bianchi_ae}
    \end{align}
with the right hand side vanishing in the non-compact body or perfect fluid case.   Note that this equation, which depends only on the general covariance of the matter action and on the assumption that it couple to both the metric and the {\AE}ther field, is equivalent to the one derived by applying the Riemannian Bianchi identity to the original \eae field equations (for discussion of this equivalence in general metric theories of gravity, see  \cite{1973PhRvD...7.3563T})

For non-spinning compact bodies we then adopt the phenomenological method pioneered by Eardley \cite{1975ApJ...196L..59E} (see also \cite{2010PhRvD..81h4060G,2013PhRvD..87j4020G}). The internal structure of the bodies depends on the {\AE}ther field because it modifies the gravitational interactions within the bodies. So the mass itself could depend on the {\AE}ther field. For a non-spinning body, the only invariant quantity available to express this dependence is $\gamma=-K_{\text{\ae} \mu}u^{\mu}$, where $u^\mu$ is the four-velocity of the body (the quantity $| K_{ae}^\mu |^2$ is not suitable since it is constrained to be unity). For a system of compact bodies we can write the matter action as a sum over bodies where the mass of each body depends on $\gamma$.
\begin{align}
    S_m=-\sum_A \int  m_A(\gamma) \, d\tau_A  \,,
    \label{eq:action0}
\end{align}
where $\tau_A$ is proper time along the world line of the $A$-th body and where $\gamma=1$ corresponds to a body which is at rest with respect to the \AE ther. In order to get the  compact body form of $T^{\mu\nu}$  and $T_{\rm ae \mu}$, we need to convert the matter action (\ref{eq:action0}) into an integral over spacetime using a four-dimensional delta-function and vary it with respect to the metric $g_{\mu\nu}$ and the {\AE}ther field $K^\mu$ respectively \cite{PW2014,2022PhRvD.106f4021T}, resulting in 
\begin{align}
T^{\alpha\beta}
& =\sum_A \frac{\delta^3\left(x-x_A\right)}{u^0 \sqrt{-g}}\biggl [\left(m_A(\gamma)-\gamma \frac{\partial m_A(\gamma)}{\partial \gamma}\right) u^\alpha u^\beta 
\nonumber \\
& \quad \quad +2 \frac{\partial m_A (\gamma)}{\partial \gamma} u^{(\alpha} K^{\beta)}_{\rm ae}\biggr ], 
\nonumber \\
T_{\rm ae \mu}&=-\sum_A \frac{\delta^3\left(x-x_A\right)}{u^0 \sqrt{-g}} \frac{\partial m_A(\gamma)}{\partial \gamma}u_\mu.
\end{align}

\noindent  We can use the above equations to write Eq.\ (\ref{eq:bianchi_ae}) as
\begin{align}
u_A^\nu \nabla_\nu &\left[ m_A(\gamma) u_A^\alpha u_{A\alpha} + m_A'(\gamma) K_{\rm ae}^\mu (g_{\mu \alpha} + u_{A,\mu} u_{A\alpha}) \right] 
\nonumber \\
&= m_A'(\gamma) u_A^\mu \nabla_\mu K_{\rm ae}^\mu \,, 
\label{eq:bianchi_ae_2}
\end{align}
where $m' \equiv \partial m_A(\gamma)/\partial \gamma$.
This in turn can be expressed in terms of coordinate time $t$ and ordinary velocity $v^\mu \equiv ( 1, dx^i/dt)$, and put in the form of a ``modified geodesic equation'' for a chosen body $A$, given by
\begin{align}
\frac{dv_A^j}{dt} &+\Gamma_{\alpha\nu}^j v_A^\alpha v_A^\nu-\Gamma_{\alpha\nu}^0 v_A^\alpha v_A^\nu v_A^j 
\nonumber \\
&=\frac{2}{u_A^0}\frac{m_A'}{m_A^*}v_A^\nu\left(g^{i\alpha}-v_A^jg^{0\alpha}\right)\partial_{[\alpha}K_{{\rm ae}\nu]} \,,
\label{eq:motion}
\end{align}
where $\Gamma_{\alpha \beta}^\gamma$ are Christoffel symbols computed from the metric, and $m_A^*\equiv m_A(\gamma)-\gamma \partial m_A(\gamma)/\partial\gamma$. These equations can also be derived directly from the effective matter action (\ref{eq:action0})  (see \cite{2022PhRvD.106f4021T}).

We now must expand the effective energy-momentum tensor and vector in a PN expansion to the required order, including the $\gamma$ dependence of the masses $m_A$. 
We first use the metric (\ref{eq:metric}), together with the constraint (\ref{eq:K0}) to expand $\gamma$ in a PN series, with the result,
\begin{equation}
\gamma = 1 + \frac{1}{2} \epsilon v^2 + \epsilon^2 \left (\frac{3}{8} v^4+ \frac{1}{2} \tilde{N} v^2
- v^j \tilde{K}_{\rm ae}^j \right) + O(\epsilon^3) \,.
\label{eq:gammaexpand} 
\end{equation}
Note that, despite appearances, the $O(\epsilon^3)$ term in $\gamma$ never actually contributes to the equations of motion at an order that concerns us.

We then expand $m_A(\gamma)$ about the asymptotic value $\gamma=1$,
\begin{align}
    m_A(\gamma) &=m_{A}+\delta\gamma\left(\frac{d m_A}{d\gamma}\right)_{\gamma =1}+\frac{1}{2}\delta\gamma^2 \left(\frac{d^2m_A}{d\gamma^2}\right)_{\gamma =1}
    \nonumber \\
    & \quad +\frac{1}{6}\delta\gamma^3\left(\frac{d^3m_A}{d\gamma^3}\right)_{\gamma =1}+... \,.
\end{align}
We define the dimensionless ``sensitivities'':
\begin{equation}
\begin{gathered}
s_A \equiv\left(\frac{d \ln m_A(\gamma)}{d \ln \gamma}\right)_{\gamma=1}\,, 
\quad s_A^{\prime} \equiv\left(\frac{d^2 \ln m_A(\gamma)}{d(\ln \gamma)^2}\right)_{\gamma=1} \,,
\nonumber \\
\quad s_A^{\prime\prime} \equiv\left(\frac{d^3 \ln m_A(\gamma)}{d(\ln \gamma)^3}\right)_{\gamma=1} \,,
\end{gathered}
\end{equation}
and write the mass of body $A$ to the required PN order as
\begin{align}
    m_A(\gamma) &=m_A\biggl[1+s_A(\gamma-1)+\frac{1}{2}a_{sA}(\gamma-1)^2
    \nonumber \\
    & \qquad +\frac{1}{6}b_{sA}(\gamma-1)^3+O((\gamma-1)^4)\biggr],
\end{align}
where we define the constant mass for each body $m_A(\gamma=1)\equiv m_A$ and 
\begin{equation}
    \begin{gathered}
        a_{sA}\equiv s_A^2-s_A+{s'_A}^2, \quad b_{sA}\equiv a'_{sA} +(s_A-2)a_{sA}.
    \end{gathered}
\end{equation}
\noindent The ``prime'' on $a'_{sA}$ denotes a derivative with respect to $\gamma$.

\subsection{Conversion of potentials to the baryon density}
\label{sec:baryondensity}

We must now convert all potentials from integrals over $\sigma, \sigma^i$, $\sigma^{i j}$, and $\sigma^j_{\rm ae}$ to integrals over a mass density $\rho^*$ defined by the constant masses $m_A$ of each body, namely
\begin{equation}
    \rho^* \equiv \sum_A m_A \delta^3 (\bm{x} - \bm{x}_A ) \,.
\end{equation}
This is often called the ``baryon density'', because, in a fluid context, it satisfies a continuity equation reflecting the conservation of baryon number.
 Substituting the expressions for the metric, Eqs.\ (\ref{eq:metric}), for the change of variables, Paper I, Eqs.\ (4.7), and PN expansions for the new fields, Paper I, Eqs.\ (5.1), we obtain, to the order required for the 2.5PN equations of motion,
\begin{widetext}

\begin{subequations}
\begin{align}
\sigma & = \rho^* \biggl [ 1+\epsilon\left(\tfrac{3}{2}(1-s)v^2-G U_\sigma\right)
+\epsilon^2 \biggl \{ \tfrac{7}{8}(1-s-a_s)v^4+\tfrac{1}{2}(1-s)v^2 GU_\sigma
-2(2-c_{14})(1-s)G v^jV_\sigma^j
 \nonumber \\
&
\qquad
+\tfrac{5}{2}G^2 U_\sigma^2 -\tfrac{1}{4} N_1+\frac{3-2c_{14}}{2(2-c_{14})}B_1
+\frac{2- c_{14}}{c_1}  G s v^j V_{{\rm ae}\sigma}^j + \left ( 1- 2s - \frac{s W_L }{2c_{14}} \right ) v^j R_1^{,j} - \frac{c_1 s W_T }{2 c_{14}} v^j  X_{K{\rm ae} 1}^{,j}
 \nonumber \\
& 
\qquad 
+\frac{1}{4} \left ( 9- \frac{1}{v_L^2}  \right ) \dot{R}_1 \biggr \} -\epsilon^{5/2}\left(\tfrac{1}{4}N_{1.5}   - \frac{3-2c_{14}}{2(2-c_{14})} B_{1.5}-s v^j K^j_{\rm ae 1.5} \right) +O\left(\epsilon^3\right)\biggr ],
\label{eq:sigma_expanded}
\\
\sigma^i & = \rho^*\biggl [(1-s)v^i +\epsilon\biggl \{ \tfrac{1}{2} (1-s-a_s) v^i v^2-G (1-s)v^i U_\sigma
+(2-c_{14})G s \left ( 2V_\sigma^i   -\frac{1}{c_1} V^i_{\rm ae \sigma} \right )
\nonumber \\
& 
\qquad 
+\frac{s}{2 c_{14}}\left( W_L R_1^{,i} + c_1 W_T X_{K{\rm ae}1}^{,i} \right) \biggr \}
 +s \epsilon^{3/2} K^i_{{\rm ae} 1.5}  +O\left(\epsilon^2\right)\biggr ],
\label{eq:sigma_i_expanded}
\\
\sigma^{ij} & = \rho^*v^iv^j\left[1-s +\epsilon\left(\tfrac{1}{2}(1-s-a_s)v^2-G (1-s) U_\sigma\right)+O\left(\epsilon^2\right)\right],
\label{eq:sigma_ij_expanded}
\\
\sigma^{ii} & = \rho^* v^2 \left[(1-s)+\epsilon\left(\tfrac{1}{2}(1-s-a_s)v^2-G (1-s) U_\sigma \right)+O\left(\epsilon^2\right)\right],
\label{eq:sigma_ii_expanded}
\\
\sigma^i_{\rm ae} & = \rho^*\left[-s v^i - \epsilon\left \{ \tfrac{1}{2} a_sv^i v^2- 2 G s v^i U_\sigma
- (2-c_{14}) Gs \left (2 V^i_\sigma - \tfrac{1}{c_1}  V^i_{\rm ae \sigma} \right )
-\tfrac{s}{2 c_{14}}\left( W_L R_1^{,i} + c_1 W_T X_{K{\rm ae}1}^{,i} \right) \right \}
\right .  \nonumber \\
& \left. 
\qquad
+ s \epsilon^{3/2}K^i_{\rm ae 1.5}+ O\left(\epsilon^2\right)\right],
\label{eq:sigma_i_ae_expanded}
\end{align}
\label{eq:sigmas}
\end{subequations}
where $U_{\sigma}$, $V^i_{\sigma}$, and $V^i_{{\rm ae}\sigma}$ are defined in terms of $\sigma$ densities (Paper I, Eqs.\ (4.25) and (4.26)), and the quantities $N_1$, $B_1$, $R_1$ $X_{K{\rm ae}1}$ and $N_{1.5}$, $B_{1.5}$, $K^j_{{\rm ae}1.5}$, $X_{K{\rm ae}1.5}$ are 1PN and 1.5PN solutions for the fields and superpotentials, given in Paper I, Eqs.\ (5.4) and (5.5), respectively.  The quantities $W_L$ and $W_T$ are defined by
\begin{equation}
W_L \equiv \left (1-\frac{c_{14}}{2} \right ) \left (1-\frac{1}{v_L^2} \right ) \,, \quad W_T \equiv 1 - \frac{1}{v_T^2} \,,
\label{eq:wavespeeds1}
\end{equation}
The sensitivities $s$ and $a_s$ become $s_A$ and $a_{sA}$ when attached to a specific body $A$.

Substituting these formulas into the definitions for $U_\sigma$ and the other potentials defined in Paper I, Eqs.\ (4.25) -- (4.27), and iterating successively, we convert all such potentials into new potentials defined using $\rho^*$, plus PN corrections. For example, we find that
 \begin{align}
U_\sigma &= U+\epsilon \left(\tfrac{3}{2} \Phi_1-G\Phi_2\right)+\epsilon^2 \biggl \{ \tfrac{7}{8} \Sigma^{jj}(v^2)-\tfrac{7}{8}\Sigma (a_s v^4)
+\tfrac{1}{2}G \Sigma^{jj}( U)-2(2-c_{14}) G \Sigma^j( V^j)
\nonumber \\
&
  \quad
-\tfrac{1}{c_1} (2-c_{14})G \Sigma^j_{\rm ae}(V^j_{\rm ae})+\tfrac{1}{2}\left(5-4 c_{14}\right)G \Sigma(\Phi_1)-\left(1-c_{14}\right)G^2\Sigma(\Phi_2)+\tfrac{1}{2}(3-c_{14})G^2 \Sigma(U^2)
\nonumber \\
&  \quad
-\tfrac{1}{4}\left[2-c_{14}\left(9-v_L^{-2}\right)\right] G \Sigma(\ddot{X}) 
-\tfrac{1}{4}\left(9-v_L^{-2}\right)\left(2-c_{14}\right) G \Sigma(\dot{X}^j_{\rm ae, j}) 
+c_{14} G \Sigma^j (\dot{X}^{,j} )
 \nonumber \\
&
\quad
- (2-c_{14}) G \Sigma^j (X^k_{{\rm ae} , jk}) 
 -\tfrac{1}{2}\left(2 c_{14}+W_L \right) G \Sigma^j_{\rm ae}(\dot{X}^{,j})
+\tfrac{1}{2 c_{14}} (2-c_{14}) \left(2c_{14}+W_L-W_T\right)G \Sigma^j_{\rm ae} (X^k_{{\rm ae} , jk}) 
\biggr \} 
\nonumber \\
& \quad
 -\epsilon^{5 / 2}\left(\tfrac{1}{3} \left (4-3 c_{14} \right ) G U \dddot{\mathcal{I}}_{j j}(t)
 +\tfrac{1}{c_1 v_T} (2-c_{14})G \dot{\mathcal{I}}^j_{\rm ae}(t)V^j_{\rm ae} 
 -\tfrac{1}{3}G \Sigma(x^j)\ddot{\mathcal{I}}^j_{\rm ae}(t)
 +\left(3-2 c_{14}\right) G U \ddot{\mathcal{I}}^{kk}_{\rm ae}(t)\right)
 \nonumber \\
&  \quad +O\left(\epsilon^3 \right) \,,
\nonumber \\
V^i_\sigma &= V^i + \epsilon \biggl \{ \tfrac{1}{2}\Sigma^i(v^2)-\tfrac{1}{2}\Sigma(a_s v^i v^2) - G V^i_2
+ (2- c_{14}) G \left (2 \Sigma_{\rm ae} (V^i) + \tfrac{1}{c_1}  \Sigma_{\rm ae}(V^i_{\rm ae}) \right)
+ \tfrac{1}{2} W_L G \Sigma_{\rm ae}(\dot{X}^{,i})
\nonumber \\
& \quad 
-\tfrac{1}{2 c_{14}} (2-c_{14}) \left(W_L-W_T\right) G \Sigma_{\rm ae} (X^j_{{\rm ae},ij}) \biggr \}
+\epsilon^{3/2} \tfrac{1}{c_1 v_T} (2-c_{14}) G U_{\rm ae} \dot{\mathcal{I}}^i_{\rm ae}(t) + O\left(\epsilon^2\right)
\,,
\nonumber \\
V^i_{\rm ae \sigma}  &=  -V^i_{\rm ae} +\epsilon \biggl \{ 2 G V^i_{2 \rm ae} - \tfrac{1}{2} \Sigma(a_s v^i v^2) 
+ (2- c_{14}) G \left (2 \Sigma_{\rm ae} (V^i) + \tfrac{1}{c_1}  \Sigma_{\rm ae}(V^i_{\rm ae}) \right)
+ \tfrac{1}{2} W_L G \Sigma_{\rm ae}(\dot{X}^{,i})
\nonumber \\
& \quad 
-\tfrac{1}{2 c_{14}} (2-c_{14}) \left(W_L-W_T\right) G \Sigma_{\rm ae} (X^j_{{\rm ae},ij}) \biggr \}
+\epsilon^{3/2} \tfrac{1}{c_1 v_T} (2-c_{14})G U_{\rm ae} \dot{\mathcal{I}}^i_{\rm ae}(t) +
O\left(\epsilon^2\right) \,,
\nonumber \\
X_{\sigma}  & = X + \epsilon \left(\tfrac{3}{2}X_1 - G X_2\right) + O\left(\epsilon^2\right) \,,
\nonumber \\
X^i_{\rm ae \sigma}  & = -X^i_{\rm ae} + \epsilon \biggl \{2 G X^i_{2\rm ae}
-\tfrac{1}{2}X(a_s v^2 v^i) 
+ (2-c_{14})  G \left ( 2X_{\rm ae}(V^i) + \tfrac{1}{c_1} X_{\rm ae}(V_{\rm ae}^i) \right )
 +\tfrac{1}{2} W_L G X_{\rm ae} (\dot{X}^{,i}) 
\nonumber \\
& 
\quad
-\tfrac{1}{2 c_{14}} (2-c_{14}) \left(W_L-W_T\right) G X_{\rm ae} (X^j_{{\rm ae} , ij}) 
\biggr \}
 +\epsilon ^{3/2} \tfrac{1}{c_1 v_T} (2-c_{14}) G X_{\rm ae} \dot{\mathcal{I}}^i_{\rm ae} + O\left(\epsilon^2\right ) \,,
\label{eq:rhopotentials}
\end{align}
\end{widetext}
where the potentials shown in these expressions are defined using $\rho^*$.  Examples include
\begin{eqnarray}
U &=& \int_{\cal M} \frac{\rho^*(t,{\bf x}^\prime)}{|{\bf x}-{\bf x}^\prime | }d^3x^\prime \,,
\nonumber \\
V^j &=&  \int_{\cal M} \frac{\left (1-s^\prime \right) \rho^* (t,{\bf x}^\prime) v'^j }{|{\bf x}-{\bf x}^\prime | } d^3x^\prime \,,
\nonumber \\
V_{\rm ae}^j &=&  \int_{\cal M} \frac{s^\prime \rho^* (t,{\bf x}^\prime) v'^j }{|{\bf x}-{\bf x}^\prime | } d^3x^\prime \,.
\end{eqnarray}
In some cases we will use the same notation as before, to avoid a proliferation of hats, tildes or subscripts.
We redefine the $\Sigma$, $X$ and $Y$ potentials by
\begin{eqnarray}
\Sigma (f) &\equiv& \int_{\cal M} \frac{\rho^*(t,{\bf x}^\prime)f(t,{\bf x}^\prime)}{|{\bf x}-{\bf x}^\prime | } d^3x^\prime \,,
\nonumber \\
\Sigma^j (f) &\equiv& \int_{\cal M} \frac{(1-s')\rho^* (t,{\bf x}^\prime) v'^j f(t,{\bf
x}^\prime)}{|{\bf x}-{\bf x}^\prime | } d^3x^\prime  \,,
\nonumber \\
\Sigma^{ij} (f) &\equiv& \int_{\cal M} \frac{(1-s')\rho^* (t,{\bf x}^\prime)
v'^iv'^j f(t,{\bf x}^\prime)}{|{\bf x}-{\bf x}^\prime | } d^3x^\prime 
\,,
\nonumber \\
\Sigma_{\rm ae} (f) &\equiv& \int_{\cal M} \frac{ s' \rho^* (t,{\bf x}^\prime)  f(t,{\bf x}^\prime)}{|{\bf x}-{\bf x}^\prime | } d^3x^\prime  \,,
\nonumber \\
\Sigma^j_{\rm ae} (f) &\equiv& \int_{\cal M} \frac{ s' \rho^* (t,{\bf x}^\prime) v'^j f(t,{\bf x}^\prime)}{|{\bf x}-{\bf x}^\prime | } d^3x^\prime  \,,
\nonumber \\
X(f)  &\equiv& \int_{\cal M} {\rho^* (t,{\bf x}^\prime)f(t,{\bf
x}^\prime)}
{|{\bf x}-{\bf x}^\prime | } d^3x^\prime  \,,
\nonumber \\
Y(f) &\equiv& \int_{\cal M} {\rho^* (t,{\bf x}^\prime)f(t,{\bf x}^\prime)}
{|{\bf x}-{\bf x}^\prime |^3 } d^3x^\prime  \,,
\end{eqnarray}
and their obvious counterparts $X^j$, $X^{ij}$, $X_{\rm ae}^j$, $Y^i$, $Y^{ij}$, $Y_{\rm ae}^j$, and so on.  Non-{\AE}ther potentials with components $j$ or $ij$ have a factor $(1-s')$ in the potential, and {\AE}ther potentials have a factor of $s'$ .   With this new convention, all the potentials defined in Paper I, Eqs.\ (4.26) and (4.27) can be redefined appropriately.  Thus, for example,
\begin{align}
\Phi_1 &=  \int_{\cal M} \frac{ (1-s') \rho'^*  v'^2 }{|{\bf x}-{\bf x}^\prime | } d^3x^\prime \,,
\nonumber \\
\Phi_2 &=  \int_{\cal M}  \frac{ \rho'^*}{|{\bf x}-{\bf x}^\prime | } U({\bf x}')  d^3x'  \,,
\nonumber \\ 
V_2^j &=   \int_{\cal M} \frac{ (1-s') \rho'^* v'^j }{|{\bf x}-{\bf x}^\prime | } U({\bf x}') d^3x^\prime \,,
\nonumber \\
V_{2{\rm ae}}^j & =  \int_{\cal M} \frac{ s' \rho'^* v'^j }{|{\bf x}-{\bf x}^\prime | } U({\bf x}') d^3x^\prime \,,
\nonumber \\
\Phi_2^j &=  \int_{\cal M} \frac{  \rho'^* }{|{\bf x}-{\bf x}^\prime | } V^j({\bf x}') d^3x^\prime \,,
\nonumber \\
\Phi_{2{\rm ae}}^j &=   \int_{\cal M} \frac{  \rho'^* }{|{\bf x}-{\bf x}^\prime | } V_{\rm ae}^j ({\bf x}') d^3x^\prime \,.
\end{align} 
Note that we have introduced a minus sign in potentials involving the transition from $\sigma^i_{\rm ae} \to -s \rho^* v^i$, arising from Eq.\ (\ref{eq:sigma_i_ae_expanded}).

After numerous substitutions and iterations, and with these definitions, the modified geodesic equation of motion (\ref{eq:motion}) of a chosen compact body leads to (compare with Eqs.\ (2.24) of \cite{2002PhRvD..65j4008P})
\begin{equation}
a^i = a^i_N + \epsilon a^i_{PN} +  \epsilon^{3/2} a^i_{1.5PN} +  \epsilon^2 a^i_{2PN} +  \epsilon^{5/2} a^i_{2.5PN} \dots \,,
\label{a-orders}
\end{equation}
Through 1.5PN order, the results are
\begin{subequations}
\begin{align}
a^i_N &=\frac{1}{1-s} G U^{,i} 
\label{eq:a-N}\,,
\\
\nonumber \\
a_{PN}^i &=  \frac{3}{2}   \frac{1}{1-s}   G     \Phi_1^{,i}-\frac{1}{1-s} G^{2}    \Phi_2^{,i}-\frac{4}{1-s}   G^{2}   U^{,i} U
 \nonumber \\
& \quad
-3 G \dot{U} v^i +4\left(1-\frac{c_{14}}{2}\right) G    \dot{V}^i
 \nonumber \\
& \quad
+\frac{((1-s)(2-3 s)+a_s)}{2(1-s)^{2}}  G U^{,i} v^2 
 \nonumber \\
& \quad
-\frac{((1-s)(4-3 s)-a_s)}{(1-s)^{2}}   G  U^{,j} v^i v^j
 \nonumber \\
& \quad
 +8\left(1-\frac{c_{14}}{2}\right)   G  V^{[i,j]} v^j   
  -\frac{(2-c_{14}) s}{(1-s) c_1}   G   \dot{V}_{\rm ae}^i
\nonumber \\
& \quad 
 -\frac{2 (2-c_{14}) s}{c_1(1-s)}   G V_{\rm ae}^{[i,j]}  v^j 
\nonumber \\
& \quad 
+\frac{1}{4}  \left(2- c_{14}
+\frac{s(2-c_{14})+c_{14}}{(1-s) v_L^2}\right)   G  \ddot{X}^{,i}
\nonumber \\
& \quad   
 +\frac{1}{2  (1-s)}\left(W_L+\frac{(2-c_{14}) (W_L-W_T) s}{c_{14}}\right)   
  \nonumber \\
& \qquad  
 \times
 G {\dot X}^{j,i}_{{\rm ae},j} \,,
 \label{eq:a1PN}
 \\
 a^i_{1.5PN} &=-\frac{1}{3  (1-s)}   \left(1+\frac{3(2-c_{14}) s}{c_{1}v_T}\right)  
  G   \ddot{\cal I}^i_{\rm ae} \,,
 \label{eq:a15PN}
 \end{align}
\end{subequations}
where $v^i$, $s$, and $a_s$ refer to the chosen body and the potentials are to be evaluated at the location of the chosen body, and where $G = 2G_0/(2-c_{14})$ is the gravitational constant.

We will defer discussion of the 2PN terms to a future publication.
The 2.5PN terms are very lengthy, so we break them up into smaller groups:

\begin{widetext}

\medskip
\noindent
Terms proportional to $v^2$:
\begin{align}
a^i_{2.5PN}( v^2) &=-\frac{1}{3} G  \ddot{\cal I}^j_{\rm ae} \left ( v^2 \delta^{ij} - 4 v^i v^j \right )
                 -\frac{(a_s-s(1-s))}{6(1-s)^2} \left (1+\frac{3(2-c_{14})}{c_1 v_T} \right ) G  \ddot{\cal I}^j_{\rm ae} \left ( v^2 \delta^{ij} +2 v^i v^j \right ) \,.
\label{eq:a2.5v2}
\end{align}

\noindent
Terms proportional to $v$:
\begin{align}
a^i_{2.5PN}( v) &= (2-c_{14}) G  \biggl ( \stackrel{(4)}{{\cal I}^{ij}} + 2\dddot{\cal I}_{\rm ae}^{(ij)} \biggr ) v^j
+G \left ( x^j  {\dddot{\cal I}_{\rm ae}^{j}} - (1-2c_{14}) {\dddot{\cal I}_{\rm ae}^{jj}}
+c_{14} \stackrel{(4)}{{\cal I}^{jj}} \right ) v^i
\nonumber \\
& \quad
+\frac{2(2-c_{14}) s}{3(1-s)c_1 v_T^3} G \left ( x^{[i} \dddot{\cal I}_{\rm ae}^{j]} + \dddot{\cal I}_{\rm ae}^{[ij]}  \right ) v^j
-\frac{2a_s(2-c_{14})}{(1-s)^2 c_1 v_T} G^2  U^{,(j} {\dot{\cal I}_{\rm ae}^{k)}} \left (v^i \delta^{jk}+ v^k \delta^{ij} \right )
\nonumber \\
& \quad
- \frac{2(2-c_{14})^2(2c_1+(1-2c_1)s)}{(1-s)c_1^2 v_T} G^2 U_{\rm ae}^{,[i} \dot{\cal I}_{\rm ae}^{j]} v^j
- \frac{2(2-c_{14})(2c_{14}+(1-2c_{14})s)}{(1-s)c_1 v_T} G^2 U^{,[i} \dot{\cal I}_{\rm ae}^{j]} v^j \,.
\label{eq:a25v1}
\end{align}  

\noindent
Terms with no $v$ factor, proportional to $G$ :
\begin{align}
a^{i(1)}_{2.5PN}( v^0) &= \frac{3}{5(1-s)} \left (1-\tfrac{5}{9} \left ( c_{14} + (2-c_{14})s \right ) \right ) G  \stackrel{(5)\quad}{{\cal I}^{\langle ij \rangle }} x^j
-\frac{2}{15(1-s)} \left (1-\tfrac{5}{6}\left ( c_{14} + (2-c_{14})s \right ) \right ) G \stackrel{(5)\quad}{{\cal I}^{ ijj }}
\nonumber \\
& \quad
-\frac{2}{3(1-s)} \left (1-\tfrac{1}{3}\left ( c_{14} + (2-c_{14})s \right ) \right ) G \epsilon^{jik}  \stackrel{(4)\quad}{{\cal J}^{ jk }}
-\frac{\left ( c_{14} + (2-c_{14})s \right )}{9(1-s)v_L^2} G  \stackrel{(5)\,}{{\cal I}^{ jj}} x^i
\nonumber \\
& \quad
-\frac{1}{30(1-s)} \left \{ 1 + \frac{(2-c_{14}) s W_T}{c_1 v_T} - \frac{\left ( c_{14} + (2-c_{14})s \right ) W_L}{(2-c_{14})} \left (1 + \frac{(2-c_{14})}{v_T} \right ) \right \} 
\nonumber \\
& \quad
\quad \times \left ( \delta^{jk}\delta^{im} + 2\delta^{ij}\delta^{km} \right ) G \stackrel{(4)\,}{{\cal I}_{\rm ae}^{m}}  x^j x^k  
-\frac{(2-c_{14}) s}{6 (1-s) c_1 v_T^3} G  \stackrel{(4)\,}{{\cal I}_{\rm ae}^{i}} r^2
\nonumber \\
& \quad
-\frac{1}{9(1-s)} \left \{ 3 - \frac{(2-c_{14}) s W_T}{c_1 v_T} - \frac{\left ( c_{14} + (2-c_{14})s \right ) W_L}{(2-c_{14})} \left (3 - \frac{(2-c_{14})}{v_T} \right ) \right \}  G \stackrel{(4)\,}{{\cal I}_{\rm ae}^{jj}} x^i   
\nonumber \\
& \quad
+\frac{2}{3} (2-c_{14}) G \stackrel{(4)\,}{{\cal I}_{\rm ae}^{(ij)}} x^j  
+\frac{1}{3} \frac{ (2-c_{14}) s}{(1-s) c_1 v_T^3} G \stackrel{(4)\,}{{\cal I}_{\rm ae}^{ij}} x^i  
+ \frac{1}{9} (2-c_{14}) \left (1+ \frac{ 3 s}{2(1-s) c_1 v_T^3} \right ) G \stackrel{(4)\quad}{{\cal I}_{\rm ae}^{ijj}}
\nonumber \\
& \quad
- \frac{2}{9} (2-c_{14}) G \stackrel{(4)\quad}{{\cal I}_{\rm ae}^{jij}}
- \frac{1}{6(1-s)} G \stackrel{(4)\quad}{{\cal I}_{\rm ae}^{j(ij)}} \,.
\label{eq:a25novG1}
\end{align}

\noindent
Terms with no $v$ factor, proportional to $G^2$ :
\begin{align}
a^{i(2)}_{2.5PN}( v^0) &= \frac{G^2}{1-s} \biggl \{ \left ( (2-c_{14}) U^{,j} \delta^{ik} + \tfrac{1}{3} (4+3c_{14}) U^{,i} \delta^{jk} - \tfrac{1}{2}  (2-c_{14}) X^{,ijk} \right )  \dddot{\cal I}^{jk}
\nonumber \\
& \quad
+2  (2-c_{14}) U^{,j} \ddot{\cal I}_{\rm ae}^{(ij)}
+ \frac{(2-c_{14})^2}{c_1^2 v_T}\left  (2c_1+(1-2c_1)s \right ) U_{\rm ae}  \ddot{\cal I}_{\rm ae}^{i}
+\tfrac{8}{3} U^{,i} \ddot{\cal I}_{\rm ae}^{j} x^j
+(1+2c_{14}) U^{,i} \ddot{\cal I}_{\rm ae}^{jj} 
\nonumber \\
& \quad
+ \frac{8}{3} \left ( 1 + \frac{3(2-c_{14})}{4c_1 v_T}  \left (c_{14}+(2-c_{14})s \right ) \right ) U \ddot{\cal I}_{\rm ae}^{i}
- \frac{2(2-c_{14})}{c_1 v_T} \left  (c_{14}+(2-c_{14})s \right ) V^{j,i} \dot{\cal I}_{\rm ae}^{j}
\nonumber \\
& \quad
- \frac{(2-c_{14})}{c_1^2 v_T} \left  (c_1+(2-c_{14})s \right ) V_{\rm ae}^{j,i} \dot{\cal I}_{\rm ae}^{j}
-\frac{(2-c_{14})}{c_1 v_T} \left  (2c_{14}+(1-2c_{14})s \right ) \dot{U}\dot{\cal I}_{\rm ae}^{i}
\nonumber \\
& \quad
-\frac{(2-c_{14})^2}{c_1^2 v_T} \left  (2c_{1}+(1-2c_{1})s \right ) \dot{U}_{\rm ae} \dot{\cal I}_{\rm ae}^{i}
-\frac{(2-c_{14})}{2c_1 c_{14} v_T} \left ( c_{14} W_T - (2-c_{14})(W_L-W_T) s \right ) X_{\rm ae}^{k,kij} \dot{\cal I}_{\rm ae}^{j}
\nonumber \\
& \quad
-\frac{(2-c_{14})}{2 c_{14}^2 v_T^3} \left ( c_{14} W_L + (2-c_{14})(W_L-W_T) s \right ) \dot{X}_{\rm ae}^{,ij} \dot{\cal I}_{\rm ae}^{j}
-(2-c_{14}) X^{,ijk}  \ddot{\cal I}_{\rm ae}^{(jk)}
\nonumber \\
& \quad
- \frac{4}{3} \left (1 + \frac{3 W_L}{8 c_1 v_T}  \left  (c_{14}+(2-c_{14})s \right ) \right ) X^{,ij}  \ddot{\cal I}_{\rm ae}^{j}
-\frac{(2-c_{14})}{2 c_1 c_{14} v_T} \left ( c_{14} W_L + (2-c_{14})(W_L-W_T) s \right ) \dot{X}_{\rm ae}^{,ij} \ddot{\cal I}_{\rm ae}^{j}
\nonumber \\
& \quad
-\Sigma^{,i}(x^j) \ddot{\cal I}_{\rm ae}^{j} + \frac{2(2-c_{14})}{ c_{14} v_T^3} \left  (2c_{14}+(1-2c_{14})s \right ) \dot{U}  \dot{\cal I}_{\rm ae}^{i}
- \frac{W_L}{2 c_1 v_T}  \left  (c_{14}+ 2 (2-c_{14})s \right ) \dot{X}^{,ij} \dot{\cal I}_{\rm ae}^{j}
\biggr \} \,.
\label{eq:a2.5novG2}
\end{align}
\noindent
Expressions for the multipole moments in these equations are provided in Appendix \ref{app:moments}.
\medskip
\end{widetext}

\section{Radiation reaction in compact binaries}
\label{sec:radreaction}

We now restrict our attention to binary systems.   For the most part, the method follows procedures that have been well established for general relativity and scalar-tensor gravity, as can be reviewed in 
\cite{2002PhRvD..65j4008P,2013PhRvD..87h4070M,2014grav.book.....P}.  We will not repeat these descriptions here, but instead will highlight the additional steps that must be taken because of the presence in this theory of dipole gravitational radiation at 1.5PN order.  

\noindent 
(i)
The 1.5PN dipole term in Eq.\ (\ref{eq:a15PN}) contains two time derivatives of the {\AE}ther dipole moment ${\cal I}^j_{\rm ae}$, so in addition to substituting the Newtonian equation of motion (\ref{eq:a-N})  for the acceleration that arises, one must include the 1PN corrections of Eq.\ (\ref{eq:a1PN}), as these will produce 2.5PN terms.

\noindent
(ii)
One must include the 1PN corrections to the {\AE}ther dipole moment itself; these are derived and displayed in Appendix \ref{app:moments}.  Since these correction terms are already of 2.5PN order, one can replace accelerations with the Newtonian equation of motion.

\noindent
(iii)
The 1PN terms in the equations of motion (\ref{eq:a1PN}) contain time derivatives of velocities, so one must include the 1.5PN dipole acceleration in addition to the Newtonian acceleration.

\noindent
(iv)
The 1PN terms in the equations of motion depend on the velocities ${\bm v}_1$ and ${\bm v}_2$.  As shown in Appendix \ref{app:energy}, the transformation to the relative velocity $\bm v$ contains 1.5PN corrections that must be included.
(In GR, the analogous transformation contains 2.5PN corrections that must be included when working to 3.5PN order
\cite{2002PhRvD..65j4008P}).

As in GR, there are direct 2.5PN terms in the acceleration, as shown in Eqs.\ (\ref{eq:a2.5v2}) -- (\ref{eq:a2.5novG2}).  The potentials must be truncated to two bodies and evaluated at the location of body 1, while the multipole moments and their time derivatives must be calculated using the expressions shown in Appendix \ref{app:moments}.

Having obtained the equations of motion for body 1, one can obtain the equation for body 2 by the transformation 
$1 \rightleftharpoons 2$,
 and then the relative equation of motion by subtracting.  For the Newtonian and 1PN equations, we obtain
 \begin{align}
a^i_N &= -\frac{ Gm_0 n^i}{(1-s_1)(1-s_2)r^2}  \,,
\nonumber \\
a^i_{PN} &= 
\frac{G { m_0}  n^i}{r^{2}  \left(1- {s_1}\right)  \left(1- {s_2}\right)}  \biggl \{ \left(1+3  \eta-\tfrac{3}{2}  {\cal S}
\right .
\nonumber \\
& \qquad
\left .
-\tfrac{1}{2}  \eta  \left(9-3  \left( {s_1}+ {s_2}\right)+6   {D_1}-2   {E_1} \right)
\right .
\nonumber \\
& \qquad
\left .
+\tfrac{1}{2}  \left(1- {s_1}\right)  \left(1- {s_2}\right)   {A_{as}}\right)  v^{2}
\nonumber \\
& \qquad
-3  \eta  \left( {E_1}- {D_1}\right)   {nv}^{2}
-\frac{G   { m_0}}{r  \left(1- {s_1}\right)  \left(1- {s_2}\right)} 
\nonumber \\
& \qquad
\times \left [2  \left(2+\eta\right)- 4  {\cal S} -\eta  \left(8+4   {D_1}-3  \left( {s_1}+ {s_2}\right)\right)\right ] \biggr \}
\nonumber \\
& \quad
+\frac{G   { m_0}   v^i nv}{r^{2}  \left(1- {s_1}\right)  \left(1- {s_2}\right)}  \left [ 2  \left(2-\eta\right)-3 {\cal S}
\right .
\nonumber \\
& \qquad
\left .
 -\eta  \left(7 +2   {D_1}+2   {E_1} - 3  \left( {s_1}+ {s_2}\right) \right)
 \right .
\nonumber \\
& \qquad
\left .
-\left(1- {s_1}\right)  \left(1- {s_2}\right)   {A_{as}}\right ] \,, 
 \label{eq:aNPN}
\end{align}
where $nv \equiv {\bm n} \cdot {\bm v}$, and
\begin{equation}
m_0 \equiv m_1 (1-s_1) +m_2 (1-s_2) \,.
\end{equation}
The quantities $D_1$ and $E_1$ are given by
\begin{align}
D_1& \equiv - (2- c_{14})\left [  (1-s_1)(1-s_2) - \frac{s_1 s_2}{2c_1} \right ] \,,
\nonumber \\
E_1 &\equiv \frac{1}{2} + D_1 + \frac{(2 - c_{14})}{2 c_{14}} s_1s_2 W_T 
\nonumber \\
& \qquad
- \frac{(c_{14} + (2 - c_{14})s_1)(c_{14} + (2 - c_{14})s_2)}{4 c_{14}} W_L' \,,
\end{align}
where $W_L' = 1-1/v_L^2$.  Equations (\ref{eq:aNPN}) are in agreement with previous work \cite{2014PhRvD..89h4067Y}.
Here and for future use, we define a number of functions of masses and sensitivities for binary systems:
\begin{align}
\eta &\equiv m_1 m_2 (1-s_1)(1-s_2)/m_0^2 \,,
\nonumber \\
\Delta &\equiv  (m_1 (1-s_1) -m_2  (1-s_2))/m_0 \,,
\nonumber \\
{\cal S} &\equiv (m_1 s_2 (1-s_1) +m_2 s_1 (1-s_2))/m_0 \,,
\nonumber \\
A_{as} &\equiv   \frac{1}{m_0^3} \left [m_1^3 a_{s2}\frac{(1-s_1)^2}{(1-s_2)^2}  +m_2^3 a_{s1} \frac{(1-s_2)^2}{(1-s_1)^2} \right ] \,,
\nonumber \\
B_{as} &\equiv \frac{1}{m_0^2} \left [ m_1^2 a_{s2} \frac{(1-s_1)^2}{(1-s_2)^2}  - m_2^2 a_{s1} \frac{(1-s_2)^2}{(1-s_1)^2} \right ] \,.
\label{eq:massfunctions}
\end{align}
Note that the quantity $\cal S$ is the same as $\cal S$ defined in \cite{2007PhRvD..76h4033F,2014PhRvD..89h4067Y}.  

The 1.5PN  contribution is given by
\begin{align}
a^i_{1.5PN} &= - \frac{G^2 m_1 m_2 }{3r^3} 
\frac{(s_1-s_2)^2}{(1-s_1)^2(1-s_2)^2}
\nonumber \\
& \qquad
\times  \left [ 1 + \frac{3 (2-c_{14})}{c_1 v_T} \right ] \left ( v^i - 3 n^i nv \right ) \,,
\label{eq:a15PNF}
\end{align}
while the 2.5PN contribution is given by
\begin{align}
a^i_{2.5PN} & = \frac{G^2 m_1 m_2}{r^3} 
\nonumber \\
& \quad
\times \left [ n^i nv \biggl ( Q_1 v^2 + Q_2 nv^2 
+  \frac{Q_3 Gm_0}{r(1-s_1)(1-s_2)} \biggr )
\right .
\nonumber \\
& \qquad
 \left .
+ v^i \biggl  ( Q_4 v^2 + Q_5 nv^2 
 +  \frac{Q_6 Gm_0}{r(1-s_1)(1-s_2)} \biggr) \right ] \,,
\label{eq:a25PN}
\end{align}
where the $Q_n$ are complicated functions of the Einstein-{\AE}ther parameters, the sensitivities and the functions defined in Eqs.\ (\ref{eq:massfunctions}).

\section{Orbital energy loss}
\label{sec:energyloss}

In this Section, we use the equations of motion to compute the rate of energy loss from the binary system.  To 1PN order, the total energy is given by Eq.\ (\ref{eq:energy}) in Appendix \ref{app:energy}.  Using Eqs.\ (\ref{eq:relvel}) to convert to the relative velocity, we obtain
\begin{equation}
E = E_N + \epsilon E_{PN} \,,
\end{equation}
where
\begin{align}
E_N &= \frac{1}{2} m_0 \eta v^2 - \frac{Gm_1 m_2}{r} \,,
\nonumber \\
E_{\rm PN} & = \tfrac{3}{8} m_0 \eta \left [ (1-3\eta) - A_{as} (1-s_1)(1-s_2) \right ] v^4 
\nonumber \\
& \quad
+  \frac{G m_1 m_2}{2r} \left [ \left (3 +\eta (1+3s_1 +3s_2) - 3{\cal S} \right ) v^2+ \eta \dot{r}^2 
\right .
\nonumber \\
& \qquad
\left .   
-  (7 + 2D_1+2E_1)v^2 - (1+D_1 -E_1) \dot{r}^2 \right ]
\nonumber \\
& \quad
 + \frac{G^2 m_0 m_1 m_2}{2r^2} (1-{\cal S}) \,.
\end{align}
(Note that the 1.5PN term in Eqs.\ (\ref{eq:relvel}) acting on the PN contributions to $E$ produces a 2.5PN contribution to $E$ that can be absorbed into a redefinition of the energy, as will be discussed below.)
We now calculate $dE/dt$, inserting the 1.5PN and 2.5PN terms in the equations of motion in the lowest-order term $m_0 \eta {\bm v} \cdot {\bm a}$, and the 1.5PN terms in place of accelerations generated in $dE_{PN}/dt$, and express the result as 
\begin{align}
\frac{dE}{dt} = \epsilon^{-1} \dot{E}_{-1PN} +  \dot{E}_{0PN} \,,
\end{align}
where we adopt the usual convention of labelling the analogue of GR quadrupole radiation as $0$PN or ``Newtonian'' order, and the dipole radiation term as $-1$PN order, since it is of order $\epsilon^{-1}$ larger.
At $-1$PN order, we have
\begin{equation}
\dot{E}_{-1PN} = \frac{G^2 m_0 \eta m_1 m_2}{r^3} Q_0 (v^2 -3 \dot{r}^2) \,,
\label{eq:ENdot15PN}
\end{equation}
where
\begin{equation}
Q_0 \equiv  \frac{(s_1-s_2)^2}{(1-s_1)^2(1-s_2)^2} \left [ 1 + \frac{3 (2-c_{14})}{c_1 v_T} \right ]  \,.
\end{equation}
The expression in Eq.\ (\ref{eq:ENdot15PN}) can be simplified in the usual way by extracting a total time derivative and moving it to the right-hand side to be absorbed in a meaningless 1.5PN correction to the definition of energy.  This makes use of the general identity \cite{1995PhRvD..52.6882I}
\begin{align}
\frac{d}{dt} \left ( \frac{v^{2s} \dot{r}^p}{r^q} \right ) &= \frac{v^{2s-2} \dot{r}^{p-1}}{r^{q+1}}  \left [ p v^4 - (p+q) v^2 \dot{r}^2
 \right .
 \nonumber \\
 & \qquad
 \left .
   + p v^2 r {\bm n} \cdot {\bm a} + 2s r \dot{r}  {\bm v} \cdot {\bm a} \right ] \,.
   \label{eq:identity}
\end{align}
In Eq.\ (\ref{eq:ENdot15PN}), we use the case $(s,p,q) = (0,1,2)$ to replace $v^2 -3\dot{r}^2$ with ${\bm x} \cdot {\bm a}$.  Inserting the Newtonian acceleration for ${\bm a}$, we obtain
\begin{align}
 \dot{E}_{-1PN} &= -\frac{G^3 m_1^2 m_2^2}{3r^4} \left ( \frac{s_1-s_2}{(1-s_1)(1-s_2)} \right )^2
 \nonumber \\
 & \qquad
 \times  \left ( 1+ \frac{3(2-c_{14})}{c_1 v_T} \right ) \,.
 \label{eq:Edot15PN}
 \end{align}
However, we must also include the 1PN contributions to  ${\bm x} \cdot {\bm a}$ and add them to the 2.5PN terms that arise from substituting the 1.5PN equations of motion into $dE_{PN}/dt$.   The result for these PN correction contributions to $\dot{E}_{2.5PN}$ is of the form
\begin{align}
\dot{E}_{0PN}^{\rm corr} &= \frac{G^2 (m_1m_2)^2}{m_0 r^3} Q_0 \biggl [ A_1(1-s_1)(1-s_2) v^2 (3\dot{r}^2 - v^2 )
 \nonumber \\
 & \qquad  + \frac{Gm_0}{r} (A_2 v^2 + A_3 \dot{r}^2 ) +  \frac{A_4 G^2m_0^2}{(1-s_1)(1-s_2) r^2} \biggr ] \,,
\end{align}
where
\begin{align}
A_1 &= - \tfrac{3}{2} \left (1 - 3\eta - (1-s_1)(1-s_2) A_{as} \right ) \,,
\nonumber \\
A_2 &= \tfrac{1}{2} \left ( 8 (1-2\eta) - 9 {\cal S} + 9\eta(s_1+s_2) 
\right .
\nonumber \\
& \qquad
\left .
+ \eta(1-10 D_1 - 2E_1 ) + (1-s_1)(1-s_2) A_{as} \right ) \,,
\nonumber \\
A_3 &= -13 (1-2\eta) + 12 {\cal S} - 12 \eta(s_1+s_2) 
\nonumber \\
& \qquad
+ \eta(1+15 D_1 + E_1 ) + (1-s_1)(1-s_2) A_{as} \,,
\nonumber \\
A_4 &= -4(1 - {\cal S}) + 3 \eta (2-s-1-s_2) + 4 \eta D_1 \,.
\label{eq:Ancoeffs}
\end{align}
Using the identity (\ref{eq:identity}) with   
 $(s,p,q) = (1,1,2)$ we can substitute $v^4 = 3v^2 \dot{r}^2 - v^2 {\bm x} \cdot {\bm a}_N - 2 \dot{r} r {\bm v} \cdot {\bm a}_N$, and then with $(s,p,q) = (0,1,3)$, we can substitute $Gm_0/(1-s_1)(1-s_2)r = v^2 - 4\dot{r}^2$, to obtain
\begin{align}
\dot{E}_{0PN}^{\rm  corr} &= -\frac{G^3 (m_1m_2)^2}{ r^4} Q_0 \left [ (A_1 -A_2- A_4) v^2 
\right .
\nonumber \\
& \qquad
\left .
+ (2A_1 - A_3 +4 A_4) \dot{r}^2  \right ] \,.
\label{eq:Edotcorr}
\end{align} 
The direct contributions to $\dot{E}$ come from contracting $a^i_{2.5PN}$ with $\eta m_0 {\bm v}$, giving 
\begin{align}
\dot{E}_{0PN}^{\rm direct} &= \frac{G^2 (m_1m_2)^2}{m_0 r^3} (1-s_1)(1-s_2) \biggl [ Q_4 v^4 + Q_2 \dot{r}^4 
\nonumber \\
& \qquad
+(Q_1+Q_5) v^2 \dot{r}^2 
\nonumber \\
& \qquad
+ \frac{Gm_0}{(1-s_1)(1-s_2)r} (Q_6 v^2 + Q_3 \dot{r}^2 ) \biggr ] \,.
\end{align} 
We use the identity (\ref{eq:identity}) with $(s,p,q) = (0,3,2)$ to substitute $\dot{r}^4 =(3/5)\dot{r}^2 (v^2 + {\bm x} \cdot {\bm a}_N )$ and then use it with  $(s,p,q) = (1,1,2)$ to substitute 
$v^4 = 3v^2 \dot{r}^2 - v^2 {\bm x} \cdot {\bm a}_N - 2 \dot{r} r {\bm v} \cdot {\bm a}_N$, obtaining
\begin{align}
\dot{E}_{0PN}^{\rm direct} &= \frac{G^2 (m_1m_2)^2}{m_0 r^3}  \left [
\frac{Gm_0}{r} \biggl ((Q_4 + Q_6) v^2 
\right .
\nonumber \\
& \qquad 
\left . 
- \tfrac{1}{5} (3Q_2 -5 Q_3 - 10 Q_4) \dot{r}^2 \biggr ) 
\right .
\nonumber \\
& \qquad 
\left . 
+ \tfrac{1}{5} (1-s_1)(1-s_2)v^2 \dot{r}^2
\right .
\nonumber \\
& \qquad 
\left . 
\times  (5Q_1 +3 Q_2 + 15 Q_4 +5 Q_5) 
    \right ] \,.
    \label{eq:Edotdirect}
\end{align} 
The $Q_n$ are very complicated functions of masses, sensitivities and \eae parameters, yet by a miraculous cancellation,
$5Q_1 +3 Q_2 + 15 Q_4 +5 Q_5 = 0$.   Combining the remainder of Eq.\ (\ref{eq:Edotdirect}) with Eq.\ (\ref{eq:Edotcorr}), we obtain
\begin{align}
\dot{E}_{0PN} &=  \frac{G^3 m_1^2 m_2^2}{r^4} \left [ (Q_4 + Q_6 + Q_0 Q_8) v^2 
\right .
\nonumber \\
& \qquad 
\left . 
- \tfrac{1}{5} (3Q_2 -5 Q_3 - 10 Q_4 -5 Q_0 Q_7) \dot{r}^2 \right ] \,,
\end{align}
where $Q_7 = -(1-s_1)(1-s_2)(2A_1-A_3+4A_4)$ and $Q_8= -(1-s_1)(1-s_2)(A_1-A_2-A_4)$.   Substituting for all the coefficients $Q_n$ and $A_n$, simplifying as much as possible, and extracting terms that survive in the limit $s_1 \to 0$ and $s_2 \to 0$, we obtain
\begin{align}
 \dot{E}_{0PN} &= - \frac{8}{15} \frac{G^3 m_1^2 m_2^2}{r^4} \biggl [ (12-5c_{14}) v^2 
\nonumber \\
& \qquad 
- \left ( 11- \frac{5}{12} c_{14} \left ( 15 - W_L' \right ) \right ) \dot{r}^2 
\nonumber \\
& \qquad 
 +
 \frac{ \left ( {\cal K}_1 v^2 + {\cal K}_2 \dot{r}^2 \right )}{(1-s_1)^2(1-s_2)^2} \biggr ] \,,
\end{align}
where the coefficients ${\cal K}_i$ are given schematically by
\begin{align}
{\cal K}_a &= {\cal K}_a^{00} + {\cal K}_a^{01} W_L' 
\nonumber \\
& \qquad + \left ( 1+ \frac{3(2-c_{14})}{c_1 v_T} \right ) \left ({\cal K}_a^{10} + {\cal K}_a^{11} W_L' \right ) \,,
\label{eq:Kaparameters}
\end{align}
where $a = (1,2)$ and $W_L' = 1 - 1/v_L^2$.  Each of the ${\cal K}_a^{\alpha\beta}$ is a function of the parameters $c_1$, $c_{14}$, $W_T$, the mass functions $m_0$, $\Delta$, $\cal S$, $\eta$, and $B_{as}$ and the sensitivities $s_1$ and $s_2$.  They all vanish when $s_1 = s_2 =0$.   In addition, when $c_{14} = 0$, the result reduces to that of GR.  The expressions for the ${\cal K}_a^{\alpha\beta}$  are long and are displayed in Appendix \ref{sec:Ks}.
\medskip

\section{Discussion}
\label{sec:discussion}

\subsection{Dipole radiation reaction}
\label{sec:dipole}

Our result for the $-1$PN dipole energy loss rate is in disagreement with the published result from numerous authors
\cite{2006PhRvD..73j4012F,2007PhRvD..76h4033F,2014PhRvD..89h4067Y}.  In our case, it is simple to trace the origin of our result to the
1.5PN acceleration shown in Eq.\ (\ref{eq:a15PN}). 
That acceleration arises in the modified geodesic equation (\ref{eq:motion}) from the following combination of terms:
\begin{align}
a^j_{1.5PN} &= \frac{1}{1-s} \left [ N_{1.5}^{,j} + \frac{2}{2-c_{14}} B_{1.5}^{,j} 
\right .
\nonumber \\
& \qquad \qquad
\left .
- s \left (\dot{K}^j_{{\rm ae}1.5} + 2{K}^{[i,j]}_{{\rm ae}1.5} v^j \right )\right ] \,,
\label{eq:a15PNraw}
\end{align}
where $N_{1.5}$, $B_{1.5}$ and ${K}^j_{{\rm ae}1.5}$ are the 1.5PN contributions listed in Eq.\ (5.5) of Paper 1,
\begin{eqnarray}
N_{1.5} &=& -\frac{2}{3} G\stackrel{(3)\quad}{{\cal I}^{kk}} -\frac{4}{3} G x^k {\ddot {\cal I}}_{\rm ae}^{k}
\,,
\nonumber \\
B_{1.5} &=& -2 G \left ( 1-\frac{1}{2} c_{14} \right )\left [\stackrel{(3)\quad}{{\cal I}^{kk}}  +2 {\ddot {\cal I}}_{\rm ae}^{kk} \right ]\,,
\nonumber \\
K_{\rm ae1.5}^j &=& \frac{2G}{c_1 v_T}  \left ( 1-\frac{1}{2} c_{14} \right )  {\dot {\cal I}}_{\rm ae}^{j} \,,
\label{eq:1.5PNsol}
\end{eqnarray}
These arise from expanding the retarded solutions for the fields, displayed in Paper I, Eq.\ (4.19).  For example, the term in $N_{1.5}$ comes from the term
$-(2/3)\int_{\cal M} \dddot{\tau}^{00} (r^2 - 2{\bm x} \cdot {\bm x}' + x'^2 ) d^3 x' $.  Using the equations $\tau^{0\nu}_{,\nu} - \tau^j_{{\rm ae},j} = 0$ and $\tau^{j\nu}_{,\nu} = 0$  that are implied by harmonic gauge in the relaxed Einstein {\AE}ther equations (Paper I, Eq.\ (4.18)), we can integrate by parts, discarding surface terms, to obtain $N_{1.5}$ shown above. Notice that because of our constraint on the \eae parameter $c_3 = - c_1$ as imposed by the neutron-star merger event GW170817/GRB170817, the speed associated with this retardation is unity (see Eq.\ (\ref{eq:fieldeqsnew})).  The first term in $N_{1.5}$ does not survive the gradient in Eq.\ (\ref{eq:a15PNraw}) (as in GR), while the second term (not present in GR) survives.  The term in $B_{1.5}$ comes from $-\int_{\cal M} \dot{\tau}^{kk} d^3x'$, but it also does not survive the gradient.  The contribution to $K_{\rm ae1.5}^j $ comes from the first term in the expansion of the retarded solution, $-v_T^{-1} \int_{\cal M} \dot{\tau}_{\rm ae}^j d^3x$.  This term depends on the ``transverse {\AE}ther speed'' $v_T$ that appears in the wave equation for $K_{\rm ae}^j$ (Eq.\ (\ref{eq:fieldeqsnew})).  While it does not survive the gradients in Eq.\ (\ref{eq:a15PNraw}), it does survive the time derivative.   Combining the terms gives Eq.\ (\ref{eq:a15PN}), which then leads directly to Eq.\ (\ref{eq:Edot15PN}).

Equation (\ref{eq:Edot15PN}) is in disagreement with the formula for the dipole energy flux first obtained by Foster for weakly self-gravitating systems  \cite{2006PhRvD..73j4012F} and by Foster and others for compact bodies \cite{2007PhRvD..76h4033F,2014PhRvD..89h4067Y}.   Translating from Foster's notation to ours, and imposing the post-2017 constraint $c_+ = c_1+c_3 = 0$, we obtain from his Eqs.\ (79) and (89) an expression of the same form as Eq.\ (\ref{eq:Edot15PN}), but with our coefficient $(1 + 3(2-c_{14})/c_1 v_T)$ replaced by the coefficient
\begin{equation}
\left (\frac{2}{c_{14} v_L^3} + \frac{2(2-c_{14})}{c_1 v_T} \right ) \,.
\end{equation}
This coefficient also appears in Eq.\ (116) of \cite{2014PhRvD..89h4067Y}.  Apart from a numerical factor of 3 vs.\ 2 in the second term, the biggest difference is that Foster et al.'s coefficient depends on the longitudinal speed $v_L$, whereas ours does not.   In our formulation of the relaxed Einstein-{\AE}ther equations in terms of new field variables, Eq.\ (\ref{eq:fieldeqsnew}), there are only two relevant speeds, the speed unity in the wave equations for the metric potentials $N$, $K^j$, and $B^{jk}$, and the speed $v_T$ in the wave equation for the {\AE}ther field.  There is no wave equation involving $v_L$, although $v_L$ (usually in the form of $W_L$) appears in many other places, including the field transformations of Paper I, Eq.\ (4.7), and the field contributions to the energy-momentum pseudotensor $\tau^{\mu\nu}$ via the parameter $c_2$.   We have been unable to find a simple resolution of this disagreement.

\subsection{2.5PN radiation reaction}
\label{sec:2.5PNradreaction}

At $0$PN order in the energy loss rate, our results bear no resemblance to published results, even in the limit $s_1 \to 0$, $s_2 \to 0$.  We believe that there are a number of reasons for this.

The previous calculations are based on a linearized approximation of the field equations, in that they do not take into account the nonlinearities in the fields.  It is well known in GR that a correct calculation of gravitational waves for gravitating systems {\em requires} including those nonlinearities (see for example \cite{1980PhRvL..45.1741W}).  The fact that the final GR result, known as the quadrupole formula, has the same {\em apparent} form as would be derived for the radiation from a non-gravitating dumbbell using linearized GR is essentially a fluke, unique to general relativity.  In the relaxed Einstein equations it relies on exploiting identities that arise from the conservation equation ${\tau^{\mu\nu}}_{,\nu} = 0$, where $\tau^{\mu\nu}$ includes both matter and non-linear field contributions (see \cite{PW2014} for discussion).  This works in GR, but there is no guarantee that it works in alternative theories of gravity.    In fact, it fails in scalar-tensor theory and numerous alternative theories \cite{1977ApJ...214..826W}.  In our relaxed 
Einstein-{\AE}ther equations, the relevant identity is in fact ${\tau^{\mu\nu}}_{,\nu} = \delta^{0\mu} \tau^j_{{\rm ae},j}$.  

Because of the linearized approximation, previous calculations did not take into account the PN corrections to the dipole moment, which would contribute to the $0$PN energy flux.  All the terms in Eq.\ (\ref{eq:Kaparameters}) with the prefactor 
$(1 + 3(2-c_{14})/c_1 v_T)$ come from various PN corrections to the dipole radiation reaction, as discussed at the beginning of Sec.\ \ref{sec:radreaction}.  

Finally, in applying their expressions for energy loss to quantities such as the period decrease in binary pulsars, previous authors use the relation $\dot{P}/P = -(3/2) \dot{E}/E$.  However this is only valid at the lowest Newtonian order.  If the energy loss involves PN corrections in $\dot{E}$, then one must include PN corrections to the relation between $P$ and $E$.  This is well known in calculations of higher-order radiation damping in GR (see eg. Eq.\ (2) of \cite{1995PhRvL..74.3515B}).

On the other hand, our work and the previous calculations are conceptually very different.  We have calculated the near-zone gravitational radiation reaction effects, while previous work has calculated the far-zone energy flux using the ``Noether current'' as a device for determining the contributions of the spin decomposition of the fields.  It seems eminently reasonable that they should be equal to each other, but in fact there is limited concrete evidence for this ``energy balance'' even in GR.  In fact, in their classic 1976 paper criticizing the state of affairs in gravitational radiation theory, Ehlers et al. \cite{1976ApJ...208L..77E} argued that there was actually {\em no} direct evidence supporting this energy balance assumption.  This criticism was somewhat ingenuous, since there were already calculations of radiation reaction at 2.5PN order that gave the correct quadrupole energy loss rate (see \cite{1980ApJ...242L.129W} for a review).  Subsequently, Pati and Will \cite{2002PhRvD..65j4008P} obtained the 3.5PN radiation-reaction terms in the binary equations of motion and showed explicitly that they gave the same energy loss rate as that obtained from the energy flux to 1PN order.   But beyond that, there is no additional support for energy balance in GR.

Given the strange properties of \eae theory, many induced by the enforced constraint on the {\AE}ther field, not to mention the Lorentz symmetry violation built into the theory, could it be possible that energy balance is not valid already at the lowest dipole-radiation order, and if not, would that invalidate the theory?  Conversely could there be a fundamental flaw either in our approach to applying the ``relaxed field equations'' methodology to the theory, or in the Noether current approach to determining the energy flux?   This will be the subject of further research.

\acknowledgments

This work was supported in part by the National Science Foundation,
Grant No.\ PHY 22-07681.   We are grateful for the hospitality of  the Institut d'Astrophysique de Paris and the Illinois Center for Advanced Studies of the Universe, where part of this work was carried out.    We also wish to acknowledge useful comments  by Enrico Barausse, Diego Blas, Ted Jacobson, Nicol\'as Yunes and Jann Zosso on a draft of this paper.

\appendix

\section{Multipole moments}
\label{app:moments}

The multipole moments needed for the 2.5PN equations of motion were defined in Paper 1, Eqs.\ (4.20), and are given by
\begin{eqnarray}
{\cal I}^Q &\equiv&  \int_{\cal M} \tau^{00} x^Q d^3x \,,
\nonumber
\\
{\cal J}^{iQ} &\equiv&  \epsilon^{iab}\int_{\cal M} \tau^{0b} x^{aQ} d^3x \,,
\nonumber
\\
{\cal I}_{\rm ae}^{jQ} &\equiv& \int_{\cal M} \tau^j_{\rm ae} x^Q d^3x \,,
\label{eq:genmoment}
\end{eqnarray}
where $\tau^{\mu\nu} = (-g) T_T^{\mu\nu}+ (16\pi G_0)^{-1} \Lambda_T^{\mu\nu}$, and 
$\tau_{\rm ae}^j = T_{\rm ae}^j+ (8\pi G_0)^{-1} \Lambda_{\rm ae}^{j}$, and where $T_T^{\mu\nu}$, $T_{\rm ae}^j$, 
$\Lambda_T^{\mu\nu}$ and $\Lambda_{\rm ae}^{j}$ are defined in Paper I, Eqs.\ (4.15) and (4.16).

Because the {\AE}ther dipole moment ${\cal I}_{\rm ae}^j$ contributes at 1.5PN order, we must evaluate it to 1PN order. Substituting Eq.\ (\ref{eq:sigma_i_ae_expanded}) for the compact support part of $\tau_{\rm ae}^j$ and the appropriate potentials for the non-compact support part $\Lambda_{\rm ae}^j$ and integrating over the near zone, we obtain
\begin{align}
{\cal I}_{\rm ae}^j &= - m_1 s_1 v_1^j  - m_2 s_2 v_2^j   - \tfrac{1}{2} \left (m_1 a_{s1} v_1^2 v_1^j +m_2 a_{s2} v_2^2 v_2^j  \right )
\nonumber \\
& \quad 
+\frac{Gm_1 m_2}{r} \biggl \{  \frac{1}{2 c_{14}} \left ( 4(1-s_1)(1-s_2) c_{14}^2 
\right.
\nonumber \\
& \qquad 
\left .
+5(s_1+s_2-2s_1s_2) c_{14} + 4 s_1 s_2 \right ) (v_1^j + v_2^j) 
\nonumber \\
& \qquad 
- \frac{3}{2} (s_1 -s_2) (v_1^j -v_2^j) 
\nonumber \\
&  \qquad 
 - \frac{(c_{14} + (2 - c_{14})s_1)(c_{14} + (2 - c_{14})s_2)}{4 c_{14}} W_L'
\nonumber \\
&  \qquad \quad 
 \times  \left [ v_1^j + v_2^j - (nv_1 + nv_2)n^j \right ]
\nonumber \\
& \qquad 
- \frac{2-c_{14}}{2 c_{14}} s_1 s_2 W_T  \left [ v_1^j + v_2^j + (nv_1 + nv_2)n^j \right ] \biggr \} \,,
\end{align} 

The other multipole moments contribute at 2.5PN order, so only their lowest-order expressions are needed.  In terms of relative coordinates, they are given by:
\begin{align}
{\cal I}^{jk} &= \frac{m_1m_2}{m_0^2} \left ( m_1 (1-s_1)^2 + m_2 (1-s_2)^2 \right ) x^j x^k \,,
\nonumber \\
{\cal I}^{jkl} &= - \frac{m_1m_2}{m_0^3} \left ( m_1^2 (1-s_1)^3 - m_2^2 (1-s_2)^3 \right ) x^j x^k x^l \,,
\nonumber \\
{\cal J}^{jk} &= - \frac{m_1m_2(1-s_1) (1-s_2)}{m_0^2} 
\nonumber \\
& \qquad  
\times \left ( m_1 (1-s_1)^2 - m_2 (1-s_2)^2 \right ) j^j x^k \,,
\nonumber \\
{\cal I}_{\rm ae}^{jk} &= -\frac{m_1m_2}{m_0^2} \left ( m_1 s_2 (1-s_1)^2 + m_2 s_1 (1-s_2)^2 \right ) v^j x^k \,,
\nonumber \\
{\cal I}_{\rm ae}^{jkl} &= \frac{m_1m_2}{m_0^3} \left ( m_1^2 s_2 (1-s_1)^3 + m_2^2 s_1 (1-s_2)^3 \right ) v^j x^k x^l \,,
\end{align}
where $j^j = ({\bm x} \times {\bm v})^j$.

\section{Total energy and momentum to PN order}
\label{app:energy}

The total energy of the system to PN order can be derived by constructing the quantity
$  m_1 (1-s_1) v_1^2 +  m_2 (1-s_2) v_2^2 $, calculating its time derivative,
and, after substituting the Newtonian and PN accelerations, showing that a combination of terms has a total time derivative that vanishes.  The result is the energy, which is constant up to PN order, given by
\begin{align}
E &= \frac{1}{2}  \left ( m_1 (1-s_1) v_1^2 +  m_2 (1-s_2) v_2^2  \right ) - \frac{Gm_1 m_2}{r}
\nonumber \\
& \quad
 + \frac{3}{8} \left ( m_1 (1-s_1-a_{s1}) v_1^4 + m_2 (1-s_2-a_{s2}) v_2^4 \right ) 
 \nonumber \\
& \quad
+  \frac{Gm_1 m_2}{r} \left [ \frac{3}{2} \left ( (1-s_1) v_1^2 + (1-s_2) v_2^2  \right ) 
\right .
 \nonumber \\
& \qquad
\left .
+ (E_1 + D_1) {\bm v}_1 \cdot {\bm v}_2 + (E_1-D_1) nv_1 nv_2 
\right .
 \nonumber \\
& \qquad
\left .
+\frac{G(m_1+m_2)}{r} \right ] \,.
\label{eq:energy}
\end{align}
Another method is to evaluate the integral of $\tau^{00}$, which is the source of the wave equation for our transformed field $N$ shown in Paper 1, Eq.\ (4.13a), namely
\begin{equation}
\Box N = -16 \pi G \tau^{00} + O(\rho \epsilon^3) \,,
\end{equation}
where $\tau^{00} = (-g) T_T^{00} + (16 \pi G_0)^{-1} \Lambda^{00}$.  Recalling that $T_T^{00} = \sigma -\sigma^{jj}$, we use the compact body expressions for the $\sigma$'s  to 1PN order (Eq.\ (\ref{eq:sigmas})), and integrate that together with $\Lambda^{00}$ from Paper I, Eq.\ (4.16a) over the near zone.  Although the integrand is quite complicated, the actual integral simplifies dramatically, so that the result is 
\begin{equation}
\int_{\cal M} \tau^{00} d^3x = m_1 + m_2 + E \,,
\end{equation}
with $E$ given by Eq.\ (\ref{eq:energy}).  This, as expected, is the total mass-energy of the system including the rest-masses of the bodies.

In a similar manner, one can obtain an expression for the conserved total momentum of the system.  By constructing the quantity $m_1 (1-s_1) v_1^j + m_2 (1-s_2) v_2^j$ and calculating its time derivative, one can find a quantity that is constant in time, through 1.5PN order, given by
\begin{align}
P^j &= m_1(1-s_1) v_1^j \left (1+\tfrac{1}{2} v_1^2 \right ) + m_2(1-s_2) v_2^j \left (1+\tfrac{1}{2} v_2^2 \right ) 
 \nonumber \\
  & \quad 
    - \frac{1}{2} \left (m_1 a_{s1} v_1^2 v_1^j + m_2 a_{s2} v_2^2 v_2^j \right ) 
  \nonumber \\
  & \quad 
  + \frac{Gm_1 m_2}{r} \left [ 3 \left ( (1-s_1) v_1^j + (1-s_2) v_2^j \right ) 
  \right .
 \nonumber \\
& \quad
\left .
+(D_1+E_1) (v_1^j + v_2^j)
      + (D_1-E_1) (nv_1 + nv_2) n^j \right ]
 \nonumber \\
  & \quad  
  + \tfrac{1}{3} \left  [ \bar{m} + \frac{3(2-c_{14})}{c_1 v_T} (m_1 s_1 + m_2 s_2 ) \right ] G \dot{\cal I}_{\rm ae}^j \,,
\end{align}
where ${\bar m} = m_1 +m_2$.
Setting $P^j = 0$, and defining the relative velocity $v^j  \equiv v_1^j - v_2^j$, we can obtain the transformation between the individual velocities and $v^j$, given by
\begin{align}
v_1^j &= \frac{m_2 (1-s_2)}{m_0} v^j + \delta v^j_{PN} + \delta v^j_{1.5PN} \,,
\nonumber \\
v_2^j &=- \frac{m_1 (1-s_1)}{m_0} v^j + \delta v^j_{PN} + \delta v^j_{1.5PN} \,,
\label{eq:relvel}
\end{align}
where 
\begin{align}
\delta v^j_{PN} &= \frac{1}{2} \eta  ( \Delta - C_{as} ) v^2 v^j 
 \nonumber \\
  & \quad
    + \frac{Gm_1 m_2}{m_0 r} \biggl [  \biggl (3(1-s_1)(1-s_2)\frac{(m_1-m_2)}{m_0} 
\nonumber \\
& \qquad
  + \Delta (D_1+E_1) \biggr ) v^j +\Delta (D_1-E_1) \dot{r} n^j  \biggr ] \,,
\end{align}
and 
\begin{align}
\delta v^j_{1.5PN} &= - \frac{1}{3m_0} \biggl [ (m_1+m_2 ) 
\nonumber \\
& \qquad
+ \frac{3(2-c_{14})}{c_1 v_T} (m_1 s_1 + m_2 s_2 ) \biggr ] G \dot{\cal I}_{\rm ae} \,,
\end{align}
where
\begin{equation}
C_{as} \equiv \frac{1}{m_0^2} \left [ m_1^2 a_{s2} \frac{(1-s_1)^2}{(1-s_2)}  - m_2^2 a_{s1} \frac{(1-s_2)^2}{(1-s_1)} \right ]\,.
\end{equation}
The 1PN correction $\delta v^j_{PN}$ will be needed when we treat the 2PN equations of motion in a forthcoming publication; the 1.5PN correction will be needed here to convert individual velocities in the 1PN equations of motion to relative velocities, inducing 2.5PN contributions.

\begin{widetext}

\section{The coefficients ${\cal K}_a^{\alpha\beta}$}
\label{sec:Ks}

For the $v^2$ term, the coefficients are:
\begin{align}
{\cal K}_1^{00}&=
\tfrac{1}{2} \left(16-5     c_{14}  \right)    ( {s_1 }+ {s_2 }-\left( {s_1 }+ {s_2 }\right)^{2}+ {s_1 }    {s_2 } )  \left(1- {s_1 }\right)   \left(1- {s_2 }\right)
+\tfrac{1}{96} (403   \left( {s_1 }+ {s_2 }\right)^{2}-924    {s_1 }    {s_2 } )   \left(1- {s_1 }\right)   \left(1- {s_2 }\right)
\nonumber \\
& \quad
+\tfrac{1}{96} \left [15   ( {s_1 }+ {s_2 }\right)   \left(2   \left( {s_1 }- {s_2 }\right)^{2}+5    {s_1 }    {s_2 }-5    {s_1 }    {s_2 }   \left( {s_1 }+ {s_2 }\right) )+164    s_1^{2}    s_2^{2}\right)
\nonumber \\
& \quad
+\tfrac{1}{48} \left [ 71   \left( {s_1 }+ {s_2 }\right)^{2}- {s_1 }    {s_2 }   \left(93+120    {s_1 }+120    {s_2 }-80    {s_1 }    {s_2 }\right)\right ]  {W_T}
\nonumber \\
& \quad
-\left [\tfrac{5}{2}   \left(2-  c_{14}  \right)   \left(2- {s_1 }- {s_2 }\right)   \left(1- {s_1 }\right)   \left(1- {s_2 }\right)
+\tfrac{1}{96} (192-176   \left( {s_1 }+ {s_2 }\right)-45   \left( {s_1 }+ {s_2 }\right)^{2}+340    {s_1 }    {s_2 } )
\right  .
\nonumber \\
& \quad \qquad
\left .
+\tfrac{5}{6}   \left( {s_1 }+ {s_2 }-2    {s_1 }    {s_2 }\right)    {W_T}\right ]  {\cal S}
\nonumber \\
& \quad
+\left [\tfrac{5}{2}   \left(2-  c_{14}  \right)   \left(1- {s_1 }\right)   \left(1- {s_2 }\right)-\tfrac{1}{96}\left(144-85   \left( {s_1 }+ {s_2 }\right)+40    {s_1 }    {s_2 }\right)-\tfrac{31}{48}    {W_T}\right ]  {\cal S}^{2}
\nonumber \\
& \quad
+\left [ \tfrac{5}{2}   \left(2-  c_{14}  \right)   \left(1- {s_1 }\right)   \left(1- {s_2 }\right)
-\tfrac{1}{96}\left(174-85   \left( {s_1 }+ {s_2 }\right)+40    {s_1 }    {s_2 }\right)
-\tfrac{1}{16} \left(29+5   \left( {s_1 }+ {s_2 }\right)-5    {s_1 }    {s_2 }\right)    {W_T}\right ]  \left( {s_1 }- {s_2 }\right)^{2}  \eta
\nonumber \\
& \quad
-\left [ \tfrac{5}{2}   \left(2-  c_{14}  \right)   \left(1- {s_1 }- {s_2 }\right)   \left(1- {s_1 }\right)   \left(1- {s_2 }\right)
+\tfrac{1}{96}  (336-261   \left( {s_1 }+ {s_2 }\right)-45   \left( {s_1 }+ {s_2 }\right)^{2}+380    {s_1 }    {s_2 } )
\right .
\nonumber \\
& \quad \qquad
\left .
+\tfrac{1}{48}\left(71    {s_1 }+71    {s_2 }-111    {s_1 }    {s_2 }\right)    {W_T}\right ]  \left( {s_1 }- {s_2 }\right)  \Delta 
\,,
\nonumber 
\\
{\cal K}_1^{01}&=
\tfrac{1}{96}\left(3- c_1 \right)   \left [2   \left( {s_1 }^{2}- {s_1 }  {s_2 }+ {s_2 }^{2}\right)-2  {\cal S}^{2}
+\left(\left(2-  c_{14}  \right)   \left(1- {s_1 }\right)   \left(1- {s_2 }\right)-2\right)   \left( {s_1 }- {s_2 }\right)  \Delta \right ]
\nonumber \\
& \quad
+\tfrac{1}{96}\left(15   \left(2-  c_{14}  \right)   \left(1- {s_1 }\right)   \left(1- {s_2 }\right)-4   \left(3- c_1 \right)\right)  \left( {s_1 }- {s_2 }\right)^{2}   \eta
\,,
\nonumber 
\\
{\cal K}_1^{10}&=
- \tfrac{1}{48}   \left [120     c_{14}      {s_1 }    {s_2 }   \left(1- {s_1 }\right)   \left(1- {s_2 }\right)-55   \left( {s_1 }+ {s_2 }\right)^{2}-5    {s_1 }    {s_2 }   \left(24-69   \left( {s_1 }+ {s_2 }\right)+76    {s_1 }    {s_2 }\right)
\right .\nonumber \\
& \quad \qquad
\left .
+ (71   \left( {s_1 }+ {s_2 }\right)^{2}- {s_1 }    {s_2 }   \left(93+120   \left( {s_1 }+ {s_2 }\right)-110    {s_1 }    {s_2 }\right) )    {W_T}+\tfrac{60    s_1^{2}    s_2^{2}}{  c_{14}  }   \left(2- {W_T}\right)\right ]
\nonumber \\
& \quad
-\tfrac{5}{48}   \left [ (8   \left( {s_1 }+ {s_2 }\right)+3   \left( {s_1 }+ {s_2 }\right)^{2}-28 {s_1 } {s_2 } )
-8   \left( {s_1 }+ {s_2 }-2    {s_1 }    {s_2 }\right)    {W_T}\right ]  {\cal S}
\nonumber \\
& \quad
+\tfrac{5}{16}   \left [ 8  c_{14}     \left(1- {s_1 }\right)   \left(1- {s_2 }\right)-\left(12-15   \left( {s_1 }+ {s_2 }\right)+20    {s_1 }    {s_2 }\right)+\tfrac{8   }{  c_{14}  }  {s_1 }    {s_2 }
+\tfrac{1}{15c_{14}}\left(31     c_{14}  -30   \left(2-  c_{14}  \right)    {s_1 }    {s_2 }\right)    {W_T}\right ]  {\cal S}^{2}
\nonumber \\
& \quad
-\tfrac{1}{48}   \left [ 5   \left(38-15   \left( {s_1 }+ {s_2 }\right)-3    {s_1 }    {s_2 }\right)-3   \left(29+5   \left( {s_1 }+ {s_2 }\right)-10    {s_1 }    {s_2 }\right)    {W_T}+\tfrac{30}{  c_{14}  }   {s_1 }    {s_2 }  \left(1- {W_T}\right)\right ]  \left( {s_1 }- {s_2 }\right)^{2} \eta
\nonumber \\
& \quad
+\tfrac{1}{48}   \left [ 120     c_{14}     \left(1- {s_1 }\right)   \left(1- {s_2 }\right)+ 5\left( {s_1 }+ {s_2 }-8    {s_1 }    {s_2 }\right)+   \left(71    {s_1 }+71    {s_2 }-81    {s_1 }    {s_2 }\right)    {W_T}
\right . 
\nonumber \\
& \quad \qquad
\left . 
+\tfrac{60}{  c_{14}  }  {s_1 }    {s_2 }  \left(2- {W_T}\right)\right ]   \left( {s_1 }- {s_2 }\right)  \Delta
-\tfrac{5}{8}   \left(1- {s_1 }\right) \left(1- {s_2 }\right)   \left( {s_1 }- {s_2 }\right)   B_{as} \,,
\nonumber 
\\
{\cal K}_1^{11}&=\tfrac{5}{16 c_{14}} \left( c_{14} +\left(2- c_{14} \right) {s_1}\right) \left( c_{14} +\left(2- c_{14} \right) {s_2}\right) {s_1} {s_2}
+\tfrac{ 1}{48} c_1 \left({s_1}^{2}+{s_2}^{2}-{s_1} {s_2}\right)
\nonumber \\
& \quad 
-\left [ \tfrac{5}{16 c_{14}} \left( c_{14} +\left(2- c_{14} \right) {s_1}\right) \left( c_{14} +\left(2- c_{14} \right) {s_2}\right)
+\tfrac{1 }{48} c_1\right ] {\cal S}^{2}
\nonumber \\
& \quad 
-\tfrac{1}{96} \left(15 \left(2- c_{14} \right) \left(1-{s_1}\right) \left(1-{s_2}\right)+4  c_1 \right) \left({s_1}-{s_2}\right)^{2} \eta
\nonumber \\
& \quad 
-  \left [\tfrac{5}{16 c_{14}}  \left( c_{14} +\left(2- c_{14} \right) {s_1}\right) \left( c_{14} +\left(2- c_{14} \right) {s_2}\right)
- \tfrac{1}{96} c_1\left(\left(2- c_{14} \right) \left(1-{s_1}\right) \left(1-{s_2}\right)-2\right)  \right ] \left({s_1}-{s_2}\right) \Delta
\,.
\end{align}
For the $\dot{r}^2$ term, the coefficients are:
\begin{align}
{\cal K}_2^{00}&=
- \tfrac{1}{8} \left [ \left ( 52-25 c_{14} \right)   (s_1+s_2-\left(s_1+s_2\right)^{2}+s_1 s_2 ) +\tfrac{1}{3} (164 \left(s_1+s_2\right)^{2}-567 s_1 s_2 ) \right ] \left(1-s_1\right)  \left(1-s_2\right)
\nonumber \\
& \quad
+\tfrac{1}{48}\left [15 \left(s_1+s_2\right)  (35 \left(s_1-s_2\right)^{2}+27 s_1 s_2 (1- s_1 -s_2) )+1339 s_1^{2} s_2^{2}\right ]
\nonumber \\
& \quad
+\tfrac{1}{144} \left [ 101 \left(s_1+s_2\right)^{2}-s_1 s_2 \left(633-330s_1-330s_2+220s_1s_2\right) \right ] {W_T}
\nonumber \\
& \quad
+\tfrac{1}{48} \left [ 150 \left(2-c_{14} \right) \left(2-s_1-s_2\right) \left(1-s_1\right)  \left(1-s_2\right)- 72+121 \left(s_1+s_2\right)-165 \left(s_1+s_2\right)^{2}
\right .
\nonumber \\
& \quad \qquad
\left .
+400 s_1 s_2+45 s_1 s_2 \left(s_1+s_2\right)
+\tfrac{55}{72}\left(s_1+s_2-2 s_1 s_2\right ) {W_T}\right ] {\cal S}
\nonumber \\
& \quad
- \tfrac{1}{144} \left [450 \left(2-c_{14} \right) \left(1-s_1\right)  \left(1-s_2\right)+ 3 \left(224+185 \left(s_1+s_2\right)-65 s_1 s_2\right)+211 {W_T}\right ] {\cal S}^2
\nonumber \\
& \quad
-  \tfrac{1}{48}\left [ 150 \left(2-c_{14} \right) \left(1-s_1\right)  \left(1-s_2\right)+\left(749+185 \left(s_1+s_2\right)-65 s_1 s_2\right)
\right .
\nonumber \\
& \quad \qquad
\left .
+\left(119-15 \left(s_1+s_2\right)+15 s_1 s_2\right) {W_T}\right ] \left(s_1-s_2\right)^{2} \eta 
\nonumber \\
& \quad
+ \tfrac{1}{144}  \left [ 450 \left(2-c_{14} \right) \left(1-s_1-s_2\right) \left(1-s_1\right)  \left(1-s_2\right)
-636+ 276 \left(s_1+s_2\right)-165 \left(s_1+s_2\right)^{2}
\right .
\nonumber \\
& \quad \qquad
\left .
+125 s_1 s_2 + 45 s_1 s_2 \left(s_1+s_2\right))
-\left(101s_1+101s_2+9 s_1 s_2\right) {W_T} \right ] \left(s_1-s_2\right) \Delta \,,
\nonumber \\
{\cal K}_2^{01} &=
-\tfrac{5}{6} \left [ \left(5-2 c_{14} \right) ( (s_1+s_2 -\left(s_1+s_2\right)^{2}+s_1 s_2 )+\left( s_1^{2}+s_2^{2}\right)\right ] \left(1-s_1\right) \left(1-s_2\right)
\nonumber \\
& \quad
+\tfrac{\left(3-c_1 \right)}{288} \left [\left(2-c_{14} \right) \left \{\left(1-s_1\right) \left(1-s_2\right) (5 \left(s_1+s_2\right)-151 \left(s_1+s_2\right)^{2}+252 s_1 s_2 )
-111 s_1 s_2 \left(s_1+s_2-s_1^{2}-s_2^{2}\right)\right \}
\right .
\nonumber \\
& \quad
\left .
-2  ( 5 \left(s_1+s_2\right)+134 \left(s_1+s_2\right)^{2}-412 s_1 s_2 )\right ]
+\left [ 
\tfrac{5}{6} \left(2 \left(2-c_{14} \right) \left(2-s_1-s_2\right)+1\right) \left(1-s_1\right) \left(1-s_2\right)
\right .
\nonumber \\
& \quad
\left .
+\tfrac{\left(3-c_1 \right)}{288}\left(\left(2-c_{14} \right)  (40 \left(s_1+s_2\right) \left(1+s_1 s_2\right)-151 \left(s_1+s_2\right)^{2}+444 s_1 s_2 ) +10 \left(s_1+s_2-2 s_1 s_2\right)\right)\right] {\cal S}
\nonumber \\
& \quad
-\left [ \tfrac{5}{3} \left(2-c_{14} \right) \left(1-s_1\right) \left(1-s_2\right)-\tfrac{\left(3-c_1 \right)}{288} \left(\left(2-c_{14} \right) \left(262-151 \left(s_1+s_2\right)+40 s_1 s_2\right)+268\right)\right ] {\cal S}^2
\nonumber \\
& \quad
- \tfrac{1}{288} \left [525 \left(2-c_{14} \right) \left(1-s_1\right) \left(1-s_2\right)
-\left(3-c_1 \right) \left(\left(2-c_{14} \right) \left(262-151 \left(s_1+s_2\right)+40 s_1 s_2\right)+546\right)\right ]  \left(s_1-s_2\right)^{2} \eta 
\nonumber \\
& \quad
+\tfrac{1}{288}\biggl [ 240\left(2 \left(2-c_{14} \right) \left(1-s_1-s_2\right)+1\right) \left(1-s_1\right) \left(1-s_2\right)
-\left(3-c_1 \right) \left \{ \left(2-c_{14} \right) \left (\left(1-s_1\right) \left(1-s_2\right) \left(134-151 \left(s_1+s_2\right)\right)
\right .\right .
\nonumber \\
& \quad \qquad
\left .  \left . 
-111 s_1 s_2 \left(2-s_1-s_2\right)\right)-2 \left(134+5 \left(s_1+s_2\right)-10 s_1 s_2\right)\right \} \biggr ] \left(s_1-s_2\right) \Delta
\,,
\nonumber \\
{\cal K}_2^{10}&=
\tfrac{5}{48} \left [ 72 c_{14}  s_1 s_2 \left(1-s_1\right) \left(1-s_2\right)+ ( 28 \left(s_1+s_2\right)^{2}-s_1 s_2 \left(315-189 s_1-189 s_2+194 s_1 s_2\right) )
\right .
\nonumber \\
& \quad \qquad
\left .
-\tfrac{1}{15} \left \{101 \left(s_1+s_2\right)^{2}-s_1 s_2 \left(633-330s_1-330s_2+670s_1s_2\right)\right \} {W_T}+\tfrac{12}{c_{14} }s_1^{2} s_2^{2} \left(6-5 {W_T}\right) \right ]
\nonumber \\
& \quad
+\tfrac{5}{144} \left [ 3 (7 \left(s_1+s_2\right)-24 \left(s_1+s_2\right)^{2}+82 s_1 s_2 )-22 \left(s_1+s_2-2 s_1 s_2\right) {W_T}\right ]  {\cal S}
\nonumber \\
& \quad
-\tfrac{5}{48}\left [72 c_{14}  \left(1-s_1\right) \left(1-s_2\right) -\left(175-168 s_1-168s_2+180 s_1 s_2\right)+\tfrac{72}{c_{14} } s_1s_2
\right .
\nonumber \\
& \quad \qquad
\left .
-\tfrac{1}{15 c_{14} } \left(211 c_{14} +450 \left(2-c_{14} \right) s_1 s_2\right) {W_T}\right ] {\cal S}^{2}
\nonumber \\
& \quad
-\tfrac{1}{48}\biggl [ 360 c_{14}  \left(1-s_1\right) \left(1-s_2\right)-15 \left(59-56 s_1-56 s_2+61 s_1 s_2\right)
\nonumber \\
& \quad \qquad
-\left(119-15 s_1-15 s_2-180 s_1 s_2\right) {W_T}+\tfrac{390 }{c_{14} } s_1 s_2\left(1-{W_T}\right) \biggr ] \left(s_1-s_2\right)^{2} \eta 
\nonumber \\
& \quad
-\tfrac{3}{144} \left [ 360 c_{14}  \left(1-s_1\right) \left(1-s_2\right)-5 \left(14 s_1+14 s_2-9 s_1 s_2\right)
-\tfrac{1}{3} \left(101 s_1+101 s_2-441 s_1 s_2\right) {W_T}
\right .
\nonumber \\
& \quad \qquad
\left .
+\frac{60 s_1 s_2}{c_{14} } \left(6-5 {W_T}\right)\right ] \left(s_1-s_2\right) \Delta
\nonumber \\
& \quad
-\tfrac{5}{4}  \left(1-s_1\right) \left(1-s_2\right)  \left(s_1-s_2\right)   B_{as} \,,
\nonumber \\
{\cal K}_2^{11}&=
-\tfrac{5}{16c_{14}} \left(c_{14} +\left(2-c_{14} \right) s_1\right) \left(c_{14} +\left(2-c_{14} \right) s_2\right) s_1 s_2 
-\tfrac{1}{144} (5 \left(s_1+s_2\right)+134 \left(s_1+s_2\right)^{2}-412 s_1 s_2 ) c_1 
\nonumber \\
& \quad
+\tfrac{1}{288}\left(2-c_{14} \right) c_1  \left [ \left(1-s_1\right) \left(1-s_2\right) (5 \left(s_1+s_2\right)-151 \left(s_1+s_2\right)^{2}+252 s_1 s_2 )-111 \left(s_1+s_2-s_1^{2}-s_2^{2}\right) s_1 s_2\right ] 
\nonumber \\
& \quad
+\tfrac{1}{288}  c_1\left [ \left(2-c_{14} \right)  (40 \left(s_1+s_2\right) \left(1+s_1 s_2\right)-151 \left(s_1+s_2\right)^{2}+444 s_1 s_2 )+10  \left(s_1+s_2-2 s_1 s_2\right)\right ] {\cal S}
\nonumber \\
& \quad
+\tfrac{1}{288} \left [ \tfrac{90}{c_{14}} \left(c_{14} +\left(2-c_{14} \right) s_1\right) \left(c_{14} +\left(2-c_{14} \right) s_2\right)+\left(2-c_{14} \right) c_1  \left(262-151 \left(s_1+s_2\right)+40 s_1 s_2\right)+268 c_1 \right ] {\cal S}^{2}
\nonumber \\
& \quad
+\tfrac{1}{288}\left [ 45 \left(2-c_{14} \right) \left(1-s_1\right) \left(1-s_2\right)+\left(2-c_{14} \right) c_1  \left(262-151 \left(s_1+s_2\right)+40 s_1 s_2\right)+ 546 c_1 \right ] \left(s_1-s_2\right)^{2} \eta 
\nonumber \\
& \quad
+\tfrac{1}{288} \left [ \tfrac{90}{c_{14}} \left(c_{14} +\left(2-c_{14} \right) s_1\right) \left(c_{14} +\left(2-c_{14} \right) s_2\right)
+2c_1  \left(134+5 \left(s_1+s_2\right)-10 s_1 s_2\right)
\right .
\nonumber \\
& \quad \qquad
\left .
-\left(2-c_{14} \right) c_1 \left(\left(1-s_1\right) \left(1-s_2\right) \left(134-151 \left(s_1+s_2\right)\right)-111 s_1 s_2 \left(2-s_1-s_2\right)\right)
\right ] \left(s_1-s_2\right) \Delta \,.
\end{align}

\section{Errata in Paper I}
\label{sec:errata}

Paper I \cite{2023PhRvD.108l4026T} contained some unfortunate typos.  In the last equation of (5.18), the correct expression for $Y_{K{\rm ae}2}$ is
\begin{equation}
Y_{K{\rm ae}2} = - \frac{2G}{c_1} \left ( 1 - \frac{c_{14}}{2} \right ) Y^j_{\rm{ae},j} \,.
\end{equation}
In the penultimate equation of (5.19), the correct expression for $X_{K{\rm ae}2.5}$ is
\begin{equation}
c_1 X_{K{\rm ae}2.5} =  - \frac{2}{9c_1 v_T}  \left (1-\frac{c_{14}}{2} \right ) G \biggl [ c_{14} r^2 
\left ( \stackrel{(3)\quad}{{\cal I}_{\rm ae}^{kk}} - \frac{3}{5} x^j  \stackrel{(3)\quad}{{\cal I}_{\rm ae}^{j}} \right ) + 9 G(c_{14}+2c_2 )  \dot{\cal I}_{\rm ae}^{j} X_{,j} \biggr ] \,.
\end{equation}

\end{widetext}


\end{document}